\documentclass[11pt]{article}

\usepackage{amssymb}
\usepackage{latexsym} 
\usepackage{theorem} 
\usepackage{graphics}

\newtheorem{lemma}{Lemma}[section]
\newtheorem{remark}[lemma]{Remark}

\newtheorem{theorem}[lemma]{Theorem}
\newtheorem{claim}{Claim}

{\theoremstyle{break}\theorembodyfont{\rmfamily}\newtheorem{algorithm}[lemma]{Algorithm}}
{\theoremstyle{break}\theorembodyfont{\rmfamily}}

\def\display{$$\vcenter\bgroup\advance\hsize by -3em\noindent%
\ignorespaces\it\refstepcounter{equation}}
\makeatletter
\def\enddisplay{\rm\egroup\leqno(\theequation)$$\global\@ignoretrue}
\makeatother

\def\qed{\hfill $\Box$\vspace{2ex}}

\begin{document}

\title{Algorithms for square-$3PC(\cdot, \cdot)$-free Berge graphs}

\author{Fr\'ed\'eric Maffray 
    \thanks{C.N.R.S., Laboratoire
    Leibniz-IMAG, 46 Avenue F\'elix Viallet, 38031 Grenoble Cedex,
    France.  E-mail: frederic.maffray@imag.fr} \\ 
    Nicolas Trotignon
    \thanks{Universit\'e Paris I, Centre d'Economie de la Sorbonne, 106--112
    boulevard de l'H\^opital, 75647 Paris cedex 13, France,
    nicolas.trotignon@univ-paris1.fr. Partially supported by ADONET
    network, a Marie Curie training network of the European
    Community.}  \\ 
    Kristina Vu\v{s}kovi\'c 
    \thanks{School of
    Computing, University of Leeds, Leeds LS2 9JT, UK. E-mail:
    vuskovi@comp.leeds.ac.uk.  Partially supported by EPSRC grant
    EP/C518225/1.}}

\date{June 13, 2007} 

\maketitle

\begin{abstract}
We consider the class of graphs containing no odd hole, no odd
antihole, and no configuration consisting of three paths between two
nodes such that any two of the paths induce a hole, and at least two
of the paths are of length 2.  This class generalizes claw-free Berge
graphs and square-free Berge graphs.  We give a combinatorial
algorithm of complexity $O(n^{7})$ to find a clique of maximum weight
in such a graph. We also consider several subgraph-detection problems
related to this class.

\


\end{abstract}

\thispagestyle{empty}

\section{Introduction}

A graph $G$ is \emph{perfect} if every induced subgraph $G'$ of $G$
satisfies $\chi(G') = \omega(G')$, where $\chi$ denotes the chromatic
number and $\omega$ the size of a maximum clique.  We say that a graph
$G$ {\em contains} a graph $H$, if $H$ is isomorphic to an induced
subgraph of $G$.  A graph $G$ is {\em $H$-free} if it does not contain
$H$.  A {\em hole} is a chordless cycle of length at least four.  
A {\em square} is a hole of length 4.
A
graph is said to be \emph{Berge} if it does not contain an odd hole
nor the complement of an odd hole.

Berge conjectured in 1960 that a graph is Berge if and only if it is
perfect.  This was proved by Chudnovsky, Robertson, Seymour and
Thomas~\cite{chudnovsky.r.s.t:spgt} in 2002.  Later, Chudnovsky,
Cornu\'ejols, Liu, Seymour and
Vu\v{s}kovi\'c~\cite{chudnovsky.c.l.s.v:reco} gave a polynomial time
algorithm that recognizes Berge graphs.  In the 1980's, Gr\"ostchel,
Lov\'asz and Schrijver~\cite{gls}, ~\cite{grostchel.l.s:color} gave a
polynomial time algorithm that for any perfect graph computes an
optimal coloring, and a clique of maximum size.  This algorithm uses
the ellipsoid method and a polynomial time separation algorithm for a
certain class of positive semidefinite matrices related to Lov\'asz's
upper bound on the Shannon capacity of a graph \cite{lo}.  The
question remains whether these optimization problems can be solved by
purely combinatorial polynomial time algorithms, avoiding the
numerical instability of the ellipsoid method.  The aim of this paper
is to give such an algorithm for finding a clique of maximum weight in
a subclass of perfect graphs that generalizes claw-free perfect
graphs and square-free perfect graphs.

\ 

{\it $3PC(\cdot,\cdot)$'s:} A $3PC(x, y)$ is a graph induced by three
chordless paths that have the same endnodes $x$ and $y$ and such that
the union of any two of them induce a hole.  We say that a graph $G$
contains a $3PC(\cdot, \cdot)$ if it contains a $3PC(x, y)$ for some
$x, y \in V(G)$.  It is easy to see that in a $3PC(x, y)$, each of the
three paths must have length at least~2.  In literature $3PC(\cdot,
\cdot)$'s are also known as \emph{thetas} in \cite{chudnovsky.seymour:theta}.  
A {\em square-$3PC(\cdot,
\cdot)$} is a $3PC(\cdot,\cdot)$ that has at least two paths of
length 2.

In this paper we give a combinatorial algorithm, with time complexity
$O(n^{7})$, that computes a maximum weight clique in every
square-$3PC(\cdot, \cdot)$-free Berge graph.  We will show that every
square-$3PC(\cdot, \cdot)$-free Berge graph has a node whose
neighborhood has no long hole (where a {\em long hole} is a hole of
length greater than 4).  This yields a linear-size decomposition tree
into square-$3PC(\cdot ,\cdot )$-free Berge graphs that have no long
hole, and then these graphs are further decomposed into co-bipartite
graphs, resulting in the total decomposition tree of size $O(n^4)$.

Recall that there is an $O(n^9)$ recognition algorithm for the class
of Berge graphs \cite{chudnovsky.c.l.s.v:reco}.  Detecting
square-$3PC(\cdot, \cdot)$'s in a graph $G$ can be done easily: it
suffices to check, for every square $a_1, a_2, a_3, a_4, a_1$, whether
$a_1$ and $a_3$ are in the same connected component of $G\setminus
((N(a_2)\cup N(a_4)) \setminus \{a_1, a_3\})$.  This takes time
$O(n^6)$.  In Section~\ref{sec:reco}, we deal with the complexity of
several subgraph-detection problems related to this class.

\ 

A \emph{claw} is a graph on nodes $u, a, b, c$ with three edges $ua,
ub, uc$.  It is easy to see that every $3PC(\cdot, \cdot)$ contains a
claw.  So $3PC(\cdot, \cdot)$-free graphs generalize claw-free graphs.
$3PC(\cdot, \cdot)$-free Berge graphs were first studied by Aossey and
Vu\v{s}kovi\'c \cite{aossey, aossey.vuskovic:3pc} in the context of
proving the Strong Perfect Graph Conjecture for this class.  The
conjecture was proved by decomposing $3PC(\cdot, \cdot)$-free Berge
graphs into claw-free graphs using star cutsets, homogeneous pairs and
$6$-joins (a new edge cutset introduced in that paper).

Clearly square-$3PC(\cdot,\cdot)$
graphs generalize square-free graphs.  In \cite{ccv} square-free Berge
graphs are decomposed by 2-joins and star cutsets into bipartite
graphs and line graphs of bipartite graphs (hence proving the Strong
Perfect Graph Conjecture for this class).

Square-$3PC(\cdot, \cdot)$-free Berge graphs contain both claw-free
Berge graphs and square-free Berge graphs, 
and it is likely that one might be able to
obtain a similar decomposition theorem that uses star cutsets and some
of the other mentioned cutsets.  And of course all Berge graphs have
been decomposed in \cite{chudnovsky.r.s.t:spgt} (thus proving the
Strong Perfect Graph Conjecture), into basic classes by skew cutsets,
2-joins and their complements, see also \cite{maria}.

Our initial idea was to try to use the above mentioned types of
decomposition theorems to develop an algorithm for finding a maximum
weight clique in a square-$3PC(\cdot, \cdot)$-free Berge graph.
Interestingly, we did end up developing a {\em decomposition based
algorithm} for finding a maximum weight clique, but it does not use
any of the types of decomposition theorems mentioned above.

Finding a maximum weight clique in a claw-free Berge graph is not
difficult.  Indeed in such a graph $G$ the neighborhood of every node
induces a co-bipartite graph (first observed in \cite{hn}, see also
\cite{hn-perfect}), so the problem reduces to $n$ instances of
the maximum weight stable set problem in a bipartite graph, which
can be reduced to a maximum flow problem and can be done in time
$O(n^3)$, see~\cite{hn-perfect}.

For a graph $G$ let $k$ denote the number of maximal cliques in $G$,
$n$ the number of nodes in $G$ and $m$ the number of edges of $G$.
Farber \cite{farber} shows that there are $O(n^2)$ maximal cliques in
any square-free graph.  Tsukiyama, Ide, Ariyoshi and Shirakawa
\cite{tias} give an $O(nmk)$ algorithm for generating all maximal
cliques of a graph, and Chiba and Nishizeki \cite{cn} improve this
complexity to $O(\sqrt{m+n} \; mk)$.  So one can generate all the
maximal cliques of a square-free graph in time $O(\sqrt{m+n} \; mn^2)$.

For square-free Berge graphs one can obtain a slightly better algorithm
by using the following characterization obtained by Parfenoff, Roussel
and Rusu \cite{prr}: every square-free Berge graph has a node whose
neighborhood is triangulated.

\ 
 
We conclude this section by defining two more types of 3-path-configurations
($3PC$'s) and wheels. 

{\it $3PC(\Delta,\Delta)$'s:} Let $x_1, x_2, x_3, y_1, y_2, y_3$ be
six distinct nodes of $G$ such that $\{x_1, x_2, x_3\}$ and $\{y_1,
y_2, y_3\}$ both induce triangles.  A $3PC(x_1x_2x_3, y_1y_2y_3)$ is a
graph induced by three chordless paths $P_1=x_1 \cdots y_1$, $P_2=x_2
\cdots y_2$ and $P_3=x_3 \cdots y_3$, such that any two of them induce
a hole.  We say that a graph $G$ contains a $3PC(\Delta, \Delta)$ if
it contains a $3PC(x_1x_2x_3, y_1y_2y_3)$ for some $x_1, x_2, x_3,
y_1, y_2, y_3 \in V(G)$.  Such graphs are also known as prisms in
\cite{chudnovsky.r.s.t:spgt} and stretchers in \cite{eflmpr}.

{\it $3PC(\Delta,\cdot)$'s:} Let $x_1, x_2, x_3, y$ be four distinct
nodes of $G$ such that $\{ x_1, x_2, x_3\}$ induces a triangle.  We
call $3PC(x_1x_2x_3, y)$ any graph induced by three chordless paths
$P_1=x_1\cdots y$, $P_2=x_2 \cdots y$ and $P_3=x_3 \cdots y$, such
that the union of any two of them induces a hole.  Note that at least
two of three paths must have length at least~2.  We say that a graph
$G$ contains a $3PC(\Delta, \cdot)$ if it contains a $3PC(x_1x_2x_3,
y)$ for some $x_1, x_2, x_3, y \in V(G)$.  Such graphs are called
pyramids in~\cite{chudnovsky.r.s.t:spgt}.

{\it Wheels:} A {\em wheel} $(H, x)$ is a graph induced by a hole $H$
and a node $x \not\in V(H)$ that has at least three neighbors in $H$.
Node $x$ is the {\em center} of the wheel.  
A subpath of $H$, of length at least 1, whose endnodes are adjacent to
$x$, and no intermediate node is adjacent to $x$, is called a
{\em sector} of $(H,x)$. A {\em short sector} is a sector of length 1,
and a {\em long sector} is a sector of length greater than 1.
A wheel is {\em odd} if it
contains an odd number of short sectors.

It is easy to see that every odd wheel and every $3PC(\Delta, \cdot)$
contains an odd hole, so Berge graphs cannot contain these two
structures.  These facts will be used repeatedly in the proofs.

\section{Finding a maximum weight clique in a square-$3PC(\cdot, \cdot)$-free
Berge graph}

We assume that we are given a graph $G$ with a weight $f(x)$
associated with every node $x$.  The problem is to find a clique of
$G$ of maximum weight, where the weight of a subset of nodes is the
sum of the weights of its elements.  The maximum weight of a clique is
denoted by $\omega_f(G)$.  The next theorem will help us solve this
problem.  

For $x\in V(G)$, $N(x)$ denotes the set of nodes of $G$ that are adjacent to 
$x$, and $N[x]=N(x) \cup \{ x \}$.
For $A \subseteq V(G)$, $G[A]$ denotes the subgraph of $G$ induced by the
node set $A$.
$G \setminus A$ denotes the subgraph of $G$ obtainbed by removing the
node set $A$, i.e. $G \setminus A=G[V(G)\setminus A]$.

\begin{theorem}\label{thm:main}
Let $G$ be a square-$3PC(\cdot, \cdot)$-free Berge graph.  Let $x$ be a
node of $G$ such that $N(x)$ contains a long hole $H$, and let $C$ be
any connected component of $G\setminus N[x]$.  Then some node
of $H$ has no neighbor in $C$.
\end{theorem}

The proof of this theorem is long and technical, and we leave it for
Section~\ref{sec:proof}.  Here we give a corollary of
Theorem~\ref{thm:main} and show how to use it in an algorithm for the
maximum clique problem.

Let ${\cal F}$ be a class of graphs.  We say that a graph $G$ is
${\cal F}$-{\em free} if $G$ does not contain any of the graphs from
${\cal F}$.

A class ${\cal F}$ of graphs satisfies {\em property (*) w.r.t.~a
graph $G$} if the following holds: for every node $x$ of $G$ such that
$G \setminus N[x] \neq \emptyset$, and for every connected
component $C$ of $G \setminus N[x]$, if $F \in {\cal F}$ is
contained in $N(x)$, then there exists a node of $F$ that has no
neighbor in $C$.

In a graph $G$, for any node $x$, let $C_1, \ldots, C_k$ be the
components of $G\setminus N[x]$, with $|C_1|\ge |C_2|\ge
\cdots\ge |C_k|$, and let the numerical vector $(|C_1|, \ldots,
|C_k|)$ be associated with $x$.  The nodes of $G$ can thus be ordered
according to the lexicographical ordering of the numerical vectors
associated with them.  Say that a node $x$ is {\em lex-maximal} if the
associated numerical vector is lexicographically maximal over all
nodes of $G$.

\begin{theorem}\label{tcv}
Let ${\cal F}$ be a class of graphs such that for every $F \in {\cal
F}$, no node of $F$ is adjacent to all the other nodes of $F$.  If
${\cal F}$ satisfies property (*) w.r.t.~a graph $G$ and $x$ is a
lex-maximal node of $G$, then $N(x)$ is ${\cal F}$-free.
\end{theorem}
{\it Proof.} Let ${\cal F}$ be a class of graphs such that for every
$F \in {\cal F}$, no node of $F$ is adjacent to all the other nodes of
$F$.  Assume that ${\cal F}$ satisfies property (*) w.r.t.~$G$.

Let $x$ be a lex-maximal node of $G$ and suppose that $N(x)$ is not
${\cal F}$-free.  Then $G$ is not a clique, and hence, since $x$ is
lex-maximal, $G \setminus N[x] \neq \emptyset$.

Let $C_1, \ldots,C_k$ be the connected components of $G\setminus
N[x]$, with $|C_1| \geq |C_2| \geq \cdots \geq |C_k|$.  Let
$N=N(x)$ and for $i=1, \ldots,k$, let $N_i=N(x) \cap N(C_i)$.

\ 

\noindent {\bf Claim 1:} $N_1 \subseteq N_2 \subseteq \cdots \subseteq
N_k$ and for every $i=1, \ldots,k$, every node of $(N \setminus
N_i) \cup (C_{i+1} \cup \cdots \cup C_k)$ is adjacent to every node of
$N_i$.

\ 

\noindent {\em Proof of Claim 1:} We argue by induction.  First we
show that every node of $(N \setminus N_1) \cup (C_2 \cup \cdots \cup
C_k)$ is adjacent to every node of $N_1$.  Assume not and let $y \in
(N \setminus N_1) \cup (C_2 \cup \cdots \cup C_k)$ be such that it is
not adjacent to $z \in N_1$.  Clearly $y$ has no neighbor in $C_1$,
but $z$ does.  So $G \setminus N[y]$ contains a connected
component that contains $C_1 \cup z$, contradicting the choice of $x$.

Now let $i>1$ and assume that $N_1 \subseteq \cdots \subseteq N_{i-1}$
and every node of $(N \setminus N_{i-1}) \cup (C_i \cup \cdots \cup
C_k)$ is adjacent to every node of $N_{i-1}$.  Since every node of
$C_i$ is adjacent to every node of $N_{i-1}$, it follows that $N_{i-1}
\subseteq N_i$.  Suppose that there exists a node $y \in (N \setminus
N_i) \cup ( C_{i+1} \cup \cdots \cup C_k)$ that is not adjacent to a
node $z \in N_i$.  Then $z \in N_i \setminus N_{i-1}$ and $z$ has a
neighbor in $C_i$.  Also $y$ is adjacent to all nodes in $N_{i-1}$ and
no node of $C_1 \cup \cdots \cup C_i$.  So there exist connected
components of $G \setminus N[y]$, $C_1^y, \ldots,C_l^y$ such
that $C_1=C_1^y, \ldots,C_{i-1}=C_{i-1}^y$ and $C_i \cup z$ is
contained in $C_i^y$.  This contradicts the choice of $x$.  This
completes the proof of Claim 1.

\ 

Since $G[N]$ is not ${\cal F}$-free, it contains $F \in {\cal F}$.  By
property (*), a node $y$ of $F$ has no neighbor in $C_k$.  By Claim 1,
$y$ is adjacent to every node of $N_k$, and no node of $N \setminus
N_k$ has a neighbor in $C=C_1 \cup \ldots \cup C_k$.  
So (since every node of $F$ has a
non-neighbor in $F$) $F$ must contain another node $z \in N\setminus
N_k$, nonadjacent to $y$.  But then $C_1, \ldots,C_k$ are connected
components of $G \setminus N[y]$ and $z$ is contained in $(G
\setminus N[y]) \setminus C$, so $y$ contradicts the choice
of $x$.  \qed

\begin{theorem}\label{thm:neishort}
Let $G$ be a square-$3PC(\cdot, \cdot)$-free Berge graph.  Let $x$ be
a lex-maximal node in $G$.  Then the neighborhood of $x$ in $G$
contains no long hole.
\end{theorem}
{\it Proof.} Let $G$ be a square-$3PC(\cdot,\cdot)$-free Berge graph
and let $x$ be a lex-maximal node of $G$.  Let ${\cal F}$ be the set of
all long holes of $G$.  By Theorem~\ref{thm:main}, ${\cal F}$
satisfies property (*) w.r.t.~$G$.  So by Theorem~\ref{tcv}, $N(x)$ is
${\cal F}$-free, i.e. long-hole-free.  \qed

Let ${\cal F}$ be the class of square-$3PC(\cdot, \cdot)$-free Berge
graphs that contain no long hole.  Suppose that we have an algorithm
$A$ that computes a clique of maximum weight for every graph in ${\cal
F}$ in time $O(n^t)$.  Then we can compute a clique of maximum weight
for every square-$3PC(\cdot, \cdot)$-free Berge graph $G$ as follows.
By Theorem~\ref{thm:neishort}, $G$ has a node $x$ whose neighborhood
contains no long hole.  Let $G_0$ be the subgraph of $G$ induced by
$N(x)$.  So $G_0$ is in ${\cal F}$.  Clearly, since every clique of
$G$ either contains $x$ or not, we have $\omega_f(G)= \max\{f(x)+
\omega_f(G_0),\ \omega_f(G\setminus \{x\})\}$.  Thus, in order to
compute $\omega_f(G)$, we need only compute $\omega_f(G_0)$ and
$\omega_f(G\setminus \{x\})$.  The former can be done by Algorithm
$A$, and the latter can be done recursively.  Note that computing the
numerical vector associated with a node takes time $O(n^2)$, and so we
can find a lex-maximal node in time $O(n^3)$.  So we can find in time
$O(n^4)$ an ordering $x_1, \ldots, x_n$ of the nodes of $G$ such that,
for each $i=1, \ldots, n$, node $x_i$ is lex-maximal in the subgraph
induced by $x_i, \ldots, x_n$.  Thus we can compute $\omega_f(G)$ for
every square-$3PC(\cdot, \cdot)$-free Berge graph $G$ in time
$O(n^4+n^{t+1})$.  Now we describe such an algorithm $A$.  For this
purpose we will use the following definition.

\paragraph*{Full star decomposition:} For $x \in V(G)$ such that $x$
is not adjacent to every node of $G\setminus \{x\}$, let $C_1, \ldots,
C_m$ be the connected components of $G \setminus N[x]$.  Note
that $N[x]$ need not be a cutset, i.e. possibly $m=1$.  The
blocks of the {\em full star decomposition} at $x$ are the graphs
$G_0, G_1, \ldots, G_m$ defined as follows: $G_0$ is the subgraph of
$G$ induced by $N(x)$ and, for $i=1, \ldots, m$, $G_i$ is the subgraph
of $G$ induced by $C_i \cup N_i$, where $N_i$ is the set of nodes of
$N(x)$ that have a neighbor in $C_i$.

\begin{remark}\label{r1}
From the construction of the blocks of full star decomposition of $G$
it follows easily that $\omega_f(G)= \max\{f(x)+\omega_f(G_0),\
\omega_f(G_1),\ \ldots,\ \omega_f(G_m)\}$.
\end{remark}
This remark shows that the problem of finding a maximum weight clique
in $G$ can be reduced to finding a maximum weight clique in some
subgraphs of $G$.  Our algorithm for finding a maximum weight clique
in a square-$3PC(\cdot, \cdot)$-free Berge graph $G$ with no long hole
consists of the following two stages:
\begin{itemize}
\item
Stage 1: A decomposition tree $T$ is constructed, where each leaf node
is co-bipartite (i.e. complement of a bipartite graph), and for each
non-leaf node $F$, the children of $F$ in $T$ represent the blocks of
a full star decomposition of $F$.
\item
Stage 2: A maximum weight clique is computed for each of the leaf
nodes, and then the algorithm backtracks along $T$ to find a maximum
weight clique for $G$ using Remark~\ref{r1}.
\end{itemize}

Finding a maximum weight clique in a co-bipartite graph is equivalent
to finding a maximum weight stable set in a bipartite graph, and it is
well-known that this problem can be reduced to a maximum flow in a
directed network associated with the bipartite graph,
see~\cite{hn-perfect}.  
From them it is easy to deduce an algorithm
(henceforth the ``co-bipartite maximum weight clique algorithm'') that
computes a maximum weight clique in a co-bipartite graph $G$ in time
$O(n^3)$.  So if the size of the decomposition tree is polynomial,
then Stage 2 of the algorithm can be performed in polynomial time.
The key difficulty is to construct $T$ in polynomial time.  To perform
Stage 1, we decompose using a full star centered at a node contained
in an independent set of size $3$.  Note that when a Berge graph does
not contain any independent set of size $3$, then it is co-bipartite.
The following lemma implies that the decomposition tree so constructed
has polynomial size.

\begin{lemma}\label{l3} 
Let $G$ be a square-$3PC(\cdot, \cdot)$-free Berge graph that contains
no long hole, and let $G_0, $ $G_1, $ $\ldots, $ $G_m$ be the blocks
of a full star decomposition at some node $x$ of $G$.  Then
the following hold: \\
(1) No independent set of $G$ of size $2$ is contained in both $G_i$ and
$G_j$ for any $1 \leq i \neq j\leq m$.  \\
(2) No independent set of $G$ of size $3$ is contained in both $G_0$ and
$G_i$ for any $1 \leq i \leq m$.
\end{lemma}
{\em Proof.} Note that in a long-hole-free graph every $3PC(\cdot,
\cdot)$ is a square-$3PC(\cdot,\cdot)$.  To prove (1),
suppose that $\{a, b\}$ is an independent set of $G$ contained in both
$G_i$ and $G_j$ with $1\leq i\neq j \leq m$.  Then $\{a, b\} \subseteq
N(x)$, and both $a$ and $b$ have neighbors in both $C_i$ and $C_j$.
But then there exists a chordless $a, b$-path $P_i$ (resp.~$P_j)$
whose intermediate nodes are in $C_i$ (resp.~$C_j)$, and hence $P_i
\cup P_j \cup x$ induces a $3PC(a, b)$, a contradiction.

To prove (2), suppose that there exists an independent set
$\{a, b, c\}$ that is contained in both $G_0$ and $G_1$.  Then $\{a,
b, c\} \subseteq N(x)$ and every node of $\{a, b, c\}$ has a neighbor
in $C_1$.  Let $u, v, t$ be neighbors of $a, b, c$ respectively in
$C_1$.  Then there is a path $P$ in $C_1$ from $u$ to $v$.  W.l.o.g.
$u, v, P$ are chosen so that $P$ is minimal.  Then $P \cup \{x, a,
b\}$ induces a hole, and since this hole cannot be long, $u=v$.  If $t
= u$ then $\{x, a, b, c, u\}$ induces a $3PC(x, u)$.  So $t \neq
u$.  Let $Q$ be a shortest path from $t$ to $u$ in $C_1$.  W.l.o.g. no
node of $Q \setminus \{t\}$ is adjacent to $c$.  Since the graph
induced by $Q \cup \{x, a, c\}$ cannot contain a long hole, $a$ is
adjacent to $t$.  By symmetry, $b$ is also adjacent to $t$, and hence
$\{x, a, b, c, t\}$ induces a $3PC(x, t)$. 
\qed

\begin{algorithm}
  \label{alg}
\begin{description}
\item[\sc Input:] 
A square-$3PC(\cdot, \cdot)$-free Berge graph $G$ with no long hole,
and a weight function $f$ on $V(G)$.
\item[\sc Output:] 
A maximum weight clique of $G$.
\item[\sc Method:]

Step~1.  Let ${\cal L}=\{G\}$, ${\cal L}'=\emptyset$ and let $T$ be a
tree that consist of a single node $G$.

Step~2.  If ${\cal L}=\emptyset$, then go to Step~3.  Otherwise,
remove a graph $F$ from ${\cal L}$.  If $F$ does not contain an
independent set of size 3, then place $F$ in ${\cal L}'$ and return to
Step~2.  Otherwise, let $\{x, y, z\}$ be an independent set of $F$.
Decompose $F$ using the full star decomposition centered at $x$.
Place the blocks of the decomposition in ${\cal L}$, add the blocks of
the decomposition to $T$ as children of $F$, and return to Step~2.

Step~3.  For every $F \in {\cal L}'$, find a maximum weight clique of
$F$ using the co-bipartite maximum weight clique algorithm.  Note that
the leaves of $T$ are precisely the graphs in ${\cal L}'$.  Using
Remark~\ref{r1}, backtrack along $T$ to a maximum weight clique of
$G$.
\item[\sc Complexity:]
$O(n^{6})$.
\end{description}
\end{algorithm}
{\it Proof.} By the definition of Step~2, the graphs in ${\cal L}'$
(that represent the leaves of $T$) do not contain any independent set
of size $3$.  Since $G$ is Berge, these graphs are co-bipartite.  So
Step~3 correctly finds a maximum weight clique.

Now we determine the complexity of the algorithm.  Consider the tree
$T$ obtained at the end of Step~2.  We show that the number of
non-leaf nodes in $T$ is $O(n^3)$.  Let $F$ be a non-leaf
node of $T$.  Let $\{x, y, z\}$ be an independent set of $F$ from Step
2 such that $F$ is decomposed by the full star centered at $x$.  We
view $\{x, y, z\}$ as the label of $F$.  By Lemma~\ref{l3} and the
fact that $x$ is not contained in any of the blocks of decomposition
of $F$, no two non-leaf nodes of $T$ have the same label.  So the
number of non-leaf nodes of $T$ is at most $n^3$.

We now show that the number of leaf nodes of $T$ is also $O
(n^3)$.  For a node $F$ of $T$, define a measure $\tau(F)$ as
follows: if $F$ is a non-leaf node of $T$, then let $\tau(F)$ be the
number of independent sets of size 3 in $F$; if $F$ is a leaf node and
$F$ has at least three siblings in the decomposition tree, then
$\tau(F)= 1$; otherwise, $\tau(F)=0$.  Let $F$ be a non-leaf node of
$T$.  Suppose that $F$ is decomposed by a full star centered at $x$.
Let $C_1, \ldots, C_m$ be the connected components of $F \setminus
N[x]$, and let $F_0, \ldots, F_m$ be the blocks of
decomposition.  We claim that the following inequality holds: \\
\hfill {$\tau(F) \geq \tau(F_0)+\tau(F_1) + \cdots +\tau(F_m).$}
\hfill (1) \\
Indeed, by Lemma~\ref{l3}, no independent set of $F$ of size 3 is
contained in more than one block of the decomposition.  So if $m<3$,
then (1) clearly holds.  Suppose that $m \geq 3$.  To show (1), it is
enough to show that the number of independent sets of $F$ of size 3
that are not contained in any of the blocks is $\geq m+1$.  For $i=1,
\ldots, m$, let $c_i$ be a node of $C_i$.  The number of sets that
contain $x$ and two nodes from $\{c_1, \ldots, c_m\}$ is ${m\choose
2}$.  The number of sets that contain three nodes from $\{c_1, \ldots,
c_m\}$ is ${m\choose 3}$.  Note that all these sets of size 3 are
independent sets of $F$ that are not contained in any of the blocks of
the decomposition.  So the number of independent sets of size 3 of $F$
that are not contained in any of the blocks is at least ${m \choose 2}
+ {m \choose 3}$.  Since $m \geq 3$, ${m \choose 2} + {m \choose 3}
\geq m+1$.  Therefore, (1) holds.

By repeated applications of (1) we get the inequality: $\tau(G) \geq
\sum \{\tau(F)\mid F\mbox{ leaf of }T\}$.  So the number of leaves $F$
of $T$ such that $\tau(F)=1$ is $O(n^3)$.  By the definition of
$\tau$, the number of leaves $F$ of $T$ such that $\tau(F)= 0$ is at
most $3$ times the number of internal nodes of $T$.  Hence, the number
of leaves $F$ of $T$ such that $\tau (F)=0$ is $O(n^3)$.  Therefore
the number of leaves of $T$ is $O(n^3)$.

So the size of $T$ is $O(n^3)$.  When the algorithm examines a
non-leaf node of $T$, it looks for an independent set of size $3$.
Since $F$ is Berge, it suffices to check whether its complement
$\overline{F}$ is bipartite, which can be done in time $O(n^2)$ with
standard breadth-first search (and this method will produce an
independent set of size $3$ whenever $\overline{F}$ is not bipartite).
So the complexity of constructing $T$ is $O(n^5)$, and the total
complexity of Step~3 is $O(n^6)$.  Therefore, the overall complexity
of the algorithm is $O(n^6)$.  \qed

This implies that we can compute $\omega_f (G)$ for every
square-$3PC(\cdot , \cdot)$-free Berge graph $G$ in time ${\cal O} (n^7)$.
In fact this algorithm can be turned into a robust algorithm (in Spinrad's
sense \cite{spin}). We would not neet to know that the input graph $G$
to the algorithm is a square-$3PC(\cdot , \cdot )$-free Berge graph.
The algorithm would then either correctly compute $\omega_f (G)$ (in all
cases when $G$ is a square-$3PC(\cdot , \cdot )$-free Berge graph, and in
some cases when it is not) or it would identify $G$ as not being a
square-$3PC(\cdot , \cdot )$-free Berge graph. As in the algorithm we have just
given, we would start by looking for a lex-maximal vertex $x$. We would then
check whether $N(x)$ is long-hole-free. If it is not, we would
terminate the algorithm, outputing that $G$ is not a 
square-$3PC(\cdot ,\cdot )$-free Berge graph (by Theorem \ref{thm:neishort}).
In the second part of the algorithm where we decompose graphs with no long
hole, at each decomposition we would verify that (1) and (2) of 
Lemma \ref{l3} hold (which can easily be done).
If one of those conditions fails, we would again terminate the algorithm,
outputing that $G$ is not a square-$3PC(\cdot ,\cdot )$-free Berge graph.
Otherwise, we would end up with an ${\cal O} (n^3)$ decomposition tree
as before. We would now just have to check whether the leaves are
co-bipartite. If they are not, we would again terminate, outputing 
that $G$ is not a square-$3PC(\cdot ,\cdot )$-free Berge graph.

\section{Proof of Theorem~\ref{thm:main}}
\label{sec:proof}

To prove Theorem~\ref{thm:main} we prove the following stronger
result.  We {\em sign} a graph by assigning $0,1$ weights to its edges
in such a way that, for every triangle in the graph, the sum of the
weights of its edges is odd.  A graph $G$ is {\em even-signable} if
there is a signing of its edges so that for every hole in $G$, the sum
of the weights of its edges is even.  Clearly, every odd-hole-free
graph is even-signable (assign weight 1 to all its edges).  The
following theorem is an easy consequence of a theorem of Truemper
\cite{truemper}.
%
\begin{theorem}
{\em (Conforti et al.  \cite{cckv-mm,cckv-us})} A graph is
even-signable if and only if it does not contain an odd wheel nor a
$3PC(\Delta,\cdot)$.
\end{theorem}

The fact that even-signable graphs do not contain odd wheels and
$3PC(\Delta, \cdot)$'s will be used throughout the proof of the next
theorem.  

Remark: Even though Theorem~\ref{thm:main2} below implies that every
square-$3PC(\cdot, \cdot)$ even-signable graph has a node whose
neighborhood is long-hole-free, finding a largest clique in a
claw-free odd-hole-free graph (and hence in a square-$3PC(\cdot,
\cdot)$-free even-signable graph) is NP-hard.  Indeed it is proved
in~\cite{murphy} that it is NP-hard to find a largest independent set
in a graph with no cycle of length $3$, $4$, or $5$, and so it is
NP-hard to find a largest clique in a graph with no stable set of size
$3$ and no hole of length $5$.

\begin{theorem}\label{thm:main2}
Let $G$ be a square-$3PC(\cdot, \cdot)$-free even-signable graph.  Let
$x$ be a node of $G$ such that $N(x)$ contains a long hole $H$, and
let $C$ be any connected component of $G\setminus N[x]$.  Then
some node of $H$ has no neighbor in $C$.
\end{theorem}
{\it Proof.} Assume that every node of $H$ has a neighbor in $C$.  We
will show that this leads to a contradiction.  Let $n$ be the length
of $H$, and let $H=h_1h_2\cdots h_nh_1$.  Note that since $(H,x)$
cannot be an odd wheel, $H$ must be of even length, so $n \geq 6$.
For any node $u$, we denote by $N_H(u)$ the set $N(u)\cap V(H)$.
\begin{claim}\label{cl:typeu}
Every node $u$ of $C$ that has a neighbor in $H$ is one of the
following five types:
\begin{itemize}
\item 
Type~$i$, $i=1, 2, 3$: $u$ has exactly $i$ neighbors in $H$, and they
are consecutive along $H$.
\item 
Type~$4$: $u$ has exactly four neighbors $h_i, h_{i+1}, h_j, h_{j+1}$
in $H$ (that appear in this order when traversing $H$ clockwise),
where $i,j$ have different parities.
\item
Type~$5$: $u$ has exactly two neighbors $h_i$ and $h_j$ in $H$, $i$
and $j$ are of the same parity and  the two subpaths of $H$ from
$h_i$ to $h_j$ are of length greater than $2$.
\end{itemize}
\end{claim}
{\it Proof.} Let $s= |N_H(u)|$.  Note that if $u$ has three pairwise
non-adjacent neighbors $a, b, c$ in $H$, then $\{x, u, a, b, c\}$
induces a square-$3PC(x, u)$, a contradiction.  Therefore $N_H(u)$ is
covered by at most two cliques of $H$ and $s\le 4$.  If $s=1$ then $u$
is of Type~1.  Suppose $s=2$ and $u$ is not of Type~2.  Let $h_i$ and
$h_j$ be the two neighbors of $u$ in $H$.  Let $H'$ be a subpath of
$H$ whose one endnode is $h_i$ and the other is $h_j$.  If $i$ and $j$
are not of the same parity, $H'$ is of odd length greater than one and
hence $H' \cup \{ u,x\}$ induces an odd wheel with center $x$.  If
$H'$ is of length 2, then $H \cup u$ induces a square-$3PC(h_i,h_j)$.
So $u$ must be of Type~5.  If $s=3$ and $u$ is not of Type~3, then
$(H, u)$ is an odd wheel.  Suppose $s=4$ and $u$ is not of Type~4.
Then $u$ has four neighbors $h_i, h_{i+1}, h_j, h_{j+1}$ in $H$ where
$i,j$ have the same parity.  Let $H'$ be a subpath of $H$ from
$h_{i+1}$ to $h_j$.  $H'$ cannot be of length one, since then $(H,u)$
is an odd wheel.  So $H'$ is of odd length greater than one, and hence
$H'\cup \{u,x\}$ induces an odd wheel with center $x$.  This proves
the claim.  \qed

\begin{claim}\label{cl:typep}
Let $P=p_1 \cdots p_k$ be a chordless path in $C$ such that $k\ge 2$,
nodes $p_1$ and $p_k$ both have neighbors in $H$, and no node of $P
\setminus \{p_1, p_k\}$ has a neighbor in $H$.  If $N_H(p_1)
\not\subseteq N_H(p_k)$ and $N_H(p_k) \not\subseteq N_H(p_1)$, then
one of the following holds:
\begin{itemize}
\item[(i)] 
$N_H(p_1) =\{a\}$, $N_H(p_k) =\{b\}$ and either $ab$ is an edge or the
two subpaths of $H \setminus \{a,b\}$ are both of even length.
\item[(ii)] 
$N_H(p_1) =\{a, b\}$, $N_H(p_k) =\{c, d\}$, $ab$ and $cd$ are edges,
and the subpaths of $H \setminus \{a,b,c,d\}$ are of even length.
\item[(iii)] For some $i \in \{ 1, \ldots,n\}$, $N_H(p_1)=\{ h_i,
h_{i+1},h_{i+2}\}$, indices taken modulo $n$, $N(p_k) \cap \{
h_i,h_{i+1},h_{i+2}\}=h_i$, and $p_k$ is of Type~3 or 5.
\item[(iv)]
Nodes $p_1$ and $p_k$ are both of Type~5 and they have a common
neighbor in $H$.
\end{itemize} 
\end{claim}
{\it Proof.} Consider the following property $S_3$: there are three
pairwise non-adjacent nodes $h_r, h_s, h_t$ of $H$ such that $p_1$ is
adjacent to $h_r$ and $h_s$ (and thus not to $h_t$) and $p_k$ is
adjacent to $h_t$ and not to $h_r, h_s$.  Note that $S_3$ does not
hold, for otherwise $P\cup\{h_r, h_s, h_t, x\}$ induces a
square-$3PC(p_1, x)$.  By Claim~\ref{cl:typeu}, $p_1$ and $p_k$ are of
Type~1, 2, 3, 4 or 5.  This leads, up to symmetry, to the following
case analysis.

First suppose that $p_1$ is of Type~4, with $N_H(p_1)= \{h_1, h_2,
h_t, h_{t+1}\}$ with $t$ even, $4\le t\le n-2$.  If $p_k$ is of
Type~1, 2 or 5, then $S_3$ holds.  If $p_k$ is of Type~3 then either
$S_3$ holds, or $N_H(p_k)= \{h_1, h_2, h_3\}$, and in this case either
$k=2$ and $\{ p_1, p_2, h_1, h_3, h_{t+1}, x\}$ induces a
$3PC(h_1p_1p_2, x)$ or $k>2$ and $(H \setminus \{h_2\}) \cup \{ p_1,
p_k\}$ induces an odd wheel with center $p_1$.  If $p_k$ is of Type~4,
then either $S_3$ holds; or $N_H(p_k)= \{h_1, h_2, h_s, h_{s+1}\}$
with $s$ even and $4\le s\le t-2< t\le n-2$, and in this case either
$k=2$ and $\{ p_1, p_2, h_1, h_{s+1}, h_{t+1}, x\}$ induces a
$3PC(h_1p_1p_2, x)$, or $k>2$ and $\{ h_{s+1}, \ldots, h_n, h_1, p_1,
p_k\}$ induces an odd wheel with center $p_1$; or $N_H(p_k)=\{h_2,
h_3, h_{t+1}, h_{t+2}\}$, and in this case either $k=2$ and $\{x, h_1,
h_2, h_3, p_1, p_2\}$ induces an odd wheel with center $h_2$ or $k>2$
and $\{x, h_2, h_{t+1}, p_1, p_k\}$ induces a square-3PC$(h_2,
h_{t+1})$.

Now suppose that $p_1$ is of Type~3, with $N_H(p_1)= \{h_1, h_2,
h_3\}$.  If $p_k$ is of Type~1, 2 or 3, then either Property $S_3$
holds, or we have outcome (iii), or $p_k$ is adjacent to $h_4$ and
$N_H(p_k) \subseteq \{h_2, h_3, h_4\}$, and in this case $(H\setminus
\{h_2, h_3\})\cup P \cup x$ induces an odd wheel with center $x$.  If
$p_k$ is of Type~5 then either $S_3$ holds, or we have outcome (iii).

Now suppose that $p_1$ is of Type~5, with $N_H(p_1)=\{ h_1, h_t\}$ and
$t$ odd, $5\le t\le n-3$.  If $p_k$ is of Type~1 or 2 then either
$S_3$ holds, or $p_k$ is adjacent to $h_2$ and $N_H(p_k) \subseteq
\{h_1, h_2\}$, and in this case $P \cup \{ x, h_2, \ldots, h_t\}$
induces an odd wheel with center $x$.  If $p_k$ is of Type~5 then
either $S_3$ holds, or we have outcome (iv), or $N_H(p_k)=\{h_2,
h_s\}$ with $s=t\pm 1$, and in this case $P\cup\{x, h_1, h_2, h_t\}$
induces a 3PC$(xh_1h_2, p_1)$.

Finally, if $p_1, p_k$ are both of Type~1, say $N_H(p_1)=h_i$ and
$N_H(p_k)= h_j$ with $i < j$, then either we have outcome (i), or $P
\cup \{h_i, \ldots, h_j, x\}$ induces an odd wheel with center $x$.
If one of $p_1, p_k$ is of Type~1 and the other is of Type~2, then
$H\cup P$ induces a $3PC(\Delta, \cdot)$.  If $p_1, p_k$ are both of
Type~2, then either we have outcome (ii), or there is a subpath $H'=
h_i\cdots h_j$ of $H$ such that $N(p_1)\cap H'=\{h_i\}$, $N(p_k)\cap
H'= \{h_j\}$ and $H'$ is of odd length, and then $P \cup H' \cup
\{x\}$ induces an odd wheel with center $x$.  This proves the claim.
\qed

\begin{claim}\label{cl:nopyr}
$C$ does not contain a path $P=p_1 \cdots p_k$ such that either $k=1$
and $p_1$ is of Type~4 or $k\ge 2$ and $H \cup P$ induces a
$3PC(\Delta, \Delta)$.
\end{claim}
{\em Proof.} Suppose that there is such a path.  Without loss of
generality we have either $k=1$ and $N_H(p_1)=\{h_1, h_2, h_t,
h_{t+1}\}$, or $k\ge 2$, $N_H(p_1) =\{h_1, h_2\}$ and $N_H(p_k)
=\{h_t, h_{t+1}\}$.  Then by Claims \ref{cl:typeu} and \ref{cl:typep},
$t$ is even.  In particular, $h_2h_t$ and $h_{t+1}h_1$ are not edges.
Since every node of $H$ has a neighbor in $C$, there exists a
chordless path $Q=q_1 \cdots q_l$ in $C$ such that $q_1$ is adjacent
to a node of $H \setminus \{h_1, h_2, h_t, h_{t+1}\}$ and $q_l$ is
adjacent to a node of $P$.  We may assume that such paths $P$ and $Q$
are chosen so that $|V(P) \cup V(Q)|$ is minimized.  Thus no node of
$Q\setminus \{q_l\}$ has a neighbor in $P$, and the only nodes of $H$
that can have a neighbor in $Q\setminus \{q_1\}$ are $h_1, h_2, h_t,
h_{t+1}$.

First suppose that $k=1$.  W.l.o.g. $q_1$ has a neighbor $h_i$ in $h_3
\cdots h_{t-1}$.  Some node $q_j$ of $Q$ must be adjacent to one of
$h_1, h_{t+1}$, else $Q \cup \{p_1, h_1, h_{t+1}, h_i, x\}$ induces a
square-$3PC(x, p_1)$.  Let $j$ be the smallest such index; say $q_j$
is adjacent to $h_1$.  Then $q_j$ is not adjacent to $h_{t+1}$, else
$\{q_1, \ldots, q_j, x, h_1, h_i, h_{t+1}\}$ induces a square-3PC$(x,
q_j)$.  By the choice of $j$, node $q_1$ is not adjacent to $h_{t+1}$.
Then $j<l$, else $\{q_1, \ldots, q_j, p_1, h_1, h_i, h_{t+1}, x\}$
induces a 3PC$(h_1p_1q_j, x)$.  Then $q_1$ has a neighbor in $h_{t+2}
\cdots h_n$, else $\{p_1, q_1, \ldots, q_j\} \cup H \setminus \{h_2,
\ldots, h_{i-1}\}$ contains a $3PC(h_th_{t+1}p_1, h_1)$.  It follows
that $q_1$ is either of Type~4, or of Type~5 not adjacent to any of
$h_1, h_2, h_t, h_{t+1}$.  By Claim~\ref{cl:typep} applied to $Q \cup
\{p_1\}$, some node $q_s$ of $Q \setminus \{q_1\}$ has a neighbor in
$H$, and we choose the smallest such $s$.  By Claim~\ref{cl:typep}
applied to $q_1\cdots q_s$, we have $N_H(q_s) \subseteq N_H(q_1)$,
which is possible only if $q_1$ is of Type~4 with a neighbor in
$\{h_1, h_2, h_t, h_{t+1}\}$; so, up to symmetry, $N_H(q_1) = \{h_n,
h_1, h_i, h_{i+1}\}$, $2< i<t$.  Let $R$ be the shortest path from
$q_1$ to $h_{t+1}$ in the subgraph induced by $Q \cup \{p_1,
h_{t+1}\}$.  If $n>t+2$, then $R \cup \{h_{t+2}, \ldots, h_n, x\}$
induces an odd wheel with center $x$, while if $n=t+2$ then $R\cup
\{h_n, x, h_i\}$ induces a $3PC(xh_{t+1}h_{t+2}, q_1)$, a
contradiction.  Therefore we must have $k\ge 2$, and $p_1$ and $p_k$
are of Type~2.
 
Now we show that either $p_1$ or $p_k$ is the only neighbor of $q_l$
in $P$.  For suppose the contrary.  Let $a, b$ be respectively the
smallest and largest integers such that $q_l$ is adjacent to nodes
$p_a$ and $p_b$ of $P$.  So either $a\neq b$ or $1<a=b<k$.  Let $j$ be
the largest integer such that $q_j$ has a neighbor in $H$.  We can
apply Claim~\ref{cl:typep} to paths $P_a = p_1\cdots p_a q_l\cdots
q_j$ and $P_b = p_k\cdots p_b q_l\cdots q_j$.  Since $N_H(q_j)$ cannot
be a subset of both $N_H(p_1)$ and $N_H(p_k)$, this implies that
either $N_H(q_j)=\{h_1, h_2, h_t, h_{t+1}\}$ or $N_H(q_j) =\{h_i,
h_{i+1}\}$ for some $i$.  In the first case, $\{q_1, \ldots, q_j\}$
contradicts the minimality of $P\cup Q$.  So we have the second case.
If $i=1$, then $\{p_b, \ldots, p_k\}\cup Q$ contradicts the minimality
of $P\cup Q$.  If $i=t$ we have a similar contradiction.  If $i\notin
\{1, t\}$, then the parity condition of Claim~\ref{cl:typep} (ii) is
violated by one of $P_a, P_b$.  Therefore, and up to symmetry, we may
assume that $p_k$ is the only neighbor of $q_l$ in $P$.

Put $p_k=q_{l+1}$.  Let $r$ be the largest index such that a node
$q_r$ of $Q$ has a neighbor in $H\setminus \{h_t, h_{t+1}\}$.  Along
the path $q_{r+1} \cdots q_{l+1}$, let $s$ be the smallest index such
that $q_s$ has a neighbor in $H$.  By the choice of $q_r$, we have
$N_H(q_s) \subseteq \{h_t, h_{t+1}\}$, and $q_s$ is of Type~1 or 2.
W.l.o.g., $q_s$ is adjacent to $h_t$.  By Claim~\ref{cl:typep} applied
to path $q_r \cdots q_s$, we have either case (a) nodes $q_r, q_s$ are
both of Type~1, or (b) nodes $q_r, q_s$ are both of Type~2, or (c)
$N_H(q_s) \subseteq N_H(q_r)$.  More precisely: \\
In case (a), we have $N_H(q_s) = \{h_t\}$ and $N_H(q_r) = \{h_i\}$ for
some even $i\neq t$.  Suppose $t+2\le i\le n$.  So $r=1$.  If $s=l$,
then $P\cup Q\cup \{x, h_2, h_t, h_i\}$ induces a 3PC$(p_kq_lh_t, x)$.
If $s<l$ then $P\cup \{q_1, \ldots, q_s\} \cup \{h_t, \ldots, h_n,
h_1\}$ induces a 3PC$(p_kh_th_{t+1}, h_i)$.  Thus $2\le i\le t-2$.  \\
In case (b), we have $N_H(q_s)=\{h_t, h_{t+1}\}$ and $N_H(q_r)=\{h_i,
h_{i+1}\}$ for some odd $i$.  If $i=1$, then $r>1$ and $\{q_1, q_2,
\ldots, q_s\}$ contradicts the minimality of $P \cup Q$.  Thus we may
assume w.l.o.g. that $3\le i\le t-1$.  \\
In case (c), we have either case (c1) node $q_r$ is of Type~2 or 3
adjacent to $h_{t-1}$ and $h_t$, or (c2) node $q_r$ is of Type~4 with
$N_H(q_r)= \{h_i, h_{i+1}, h_t, h_{t+1}\}$ for some odd $i$ with $3\le
i\le t-3$, or (c3) node $q_r$ is of Type~4 and not adjacent to one of
$h_t, h_{t+1}$, or (c4) node $q_r$ is of Type~5 adjacent to $h_t$ and
$h_i$ for some even $i\neq t-2, t, t+2$.  In case (c3), $\{q_r,
\ldots, q_l, p_k\}$ contradicts the minimality of $P\cup Q$.  In case
(c4), suppose $t+2\le i\le n$.  If $r=l$, then $P\cup \{q_r, x, h_2,
h_t, h_i\}$ induces a 3PC$(p_kq_lh_t, x)$, while if $r<l$ then $P\cup
\{q_r\} \cup \{h_t, \ldots, h_n, h_1\}$ induces a 3PC$(p_kh_th_{t+1},
h_i)$.  Thus $2\le i\le t-2$.  \\
So we have cases (a), (b), (c1), (c2) or (c4), and in either case
there is an index $i$, with $2\le i\le t-1$, such that $q_r$ is
adjacent to $h_i$ and $N_H(q_r)\subseteq\{h_i, \ldots, h_{t+1}\}$.
Now, if $h_{t+1}$ has no neighbor in $q_r \cdots q_l$, then $P \cup (H
\setminus \{h_{i+1}, \ldots, h_t\}) \cup \{q_r, \ldots , q_l\}$
induces a $3PC(h_1h_2p_1, p_k)$.  If $h_{t+1}$ has a neighbor in $q_r
\cdots q_{l-1}$, then $P \cup (H \setminus \{h_{i+1}, \ldots, h_t\})
\cup \{q_r, \ldots, q_{l-1}\}$ contains a $3PC(h_1h_2p_1, h_{t+1})$.
So $q_l$ is the unique neighbor of $h_{t+1}$ in $q_r \cdots q_l$.  If
$i=2$, then $P \cup \{q_r, \ldots, q_l, x, h_2, h_{t+1}\}$ induces a
$3PC(h_{t+1}p_kq_l, h_2)$.  If $i>2$, then $P\cup \{q_r, \ldots, q_l,
x, h_1, h_i, h_{t+1}\}$ induces a $3PC(h_{t+1}p_kq_l, x)$.  This
proves the claim.  \qed

\

For $i=1, \ldots, n$, let $H_i$ be the set of Type~3 nodes of $C$
adjacent to $h_i, h_{i+1}, h_{i+2}$ (indices taken modulo $n$).  Note
that $H_i$ induces a clique, for if $a, b\in H_i$ are not adjacent,
then $\{a, b, x, h_i, h_{i+2}\}$ induces a square-$3PC(h_i, h_{i+2})$.

\begin{claim}\label{pyz}
Let $h_s, h_t$ be non-adjacent nodes of $H$ and $P=y \cdots z$ be a
chordless path in $C$ such that $y$ is the only neighbor of $h_s$ in
$P$ and $z$ is the only neighbor of $h_t$ in $P$.  If both $h_{s-1},
h_{s+1}$ have a neighbor in $P$, then $N_H(y)=\{h_{s-1}, h_s,
h_{s+1}\}$.
\end{claim}
{\em Proof.} We may assume up to symmetry that $s=2$ and $t\neq 4$.
Let $H'$ be the hole induced by $P\cup\{x, h_2, h_t\}$.  Since $(H',
h_3)$ cannot be an odd wheel, $(H'\setminus \{x\}) \cup \{h_3\}$ must
contain an odd number of triangles.  Let $R$ be any long sector of
$(H', h_1)$.  If $R\cup \{h_3\}$ contains an odd number of triangles
and $R$ contains at least three neighbors of $h_3$, then $R \cup
\{h_1, h_3 \}$ induces an odd wheel with center $h_3$.  If $R$
contains only two adjacent neighbors $a,b$ of $h_3$, then $R$ cannot
contain $x$ and $h_t$; and then, if $R$ does not contain $h_2$ then
$R\cup \{h_1, h_3, x\}$ induces a 3PC$(h_3ab, h_1)$, while if $R$
contains $h_2$ then $\{a, b\}=\{h_2, y\}$ and $R\cup\{h_1, h_3, x\}$
induces an odd wheel with center $h_2$.  Thus $R \cup \{h_3\}$
contains an even number of triangles for every long sector $R$ of
$(H', h_1)$.  It follows that some edge of $H'\setminus \{x\}$ is a
short sector of both $(H', h_1)$ and $(H', h_3)$, and thus some node
of $P$ is adjacent to $h_1$ and $h_3$; by Claims~\ref{cl:typeu}
and~\ref{cl:nopyr}, such a node is of Type 3 adjacent to $h_1, h_2,
h_3$ and so it can only be $y$.  This proves the claim.  \qed


\begin{claim}\label{upk}
Let $u$ be a node of $C$ that has a neighbor in $h_4 \cdots h_n$.  Let
$P=p_1\cdots p_k$ be a path of $C$ such that $p_k=u$, $p_1\in H_1$ and
no node of $P\setminus \{p_1\}$ belongs to $H_1$.  Then exactly one
node of $h_1, h_3$ has a neighbor in $P\setminus \{p_1\}$, say $h_1$
does.  Node $h_1$ must in fact have a neighbor in $P \setminus \{p_1,
p_2\}$, and so $k\ge 3$.  Furthermore, if $Q=q_1\cdots q_l$ is any
other path of $C$ such that $q_l=u$, $q_1 \in H_1$ and no node of $Q
\setminus \{q_1\}$ belongs to $H_1$, then $h_1$ has a neighbor in $Q
\setminus \{q_1\}$.  
\end{claim}
{\em Proof.} For let $h_i$ be any neighbor of $u$ in $h_4\cdots h_n$,
and let $j$ be the smallest index such that $h_i$ is adjacent to
$p_j$.  Suppose that neither $h_1$ nor $h_3$ has a neighbor in $p_2
\cdots p_j$.  If $i\notin\{4, n\}$, then $\{p_1, \ldots, p_j, h_1,
h_3, h_i, x\}$ induces a square-$3PC(p_1, x)$.  Otherwise, w.l.o.g.
$i=4$, and hence the same node set induces a $3PC(h_3h_4x, p_1)$.
Therefore $h_1$ or $h_3$ must have a neighbor in $P \setminus
\{p_1\}$.

Suppose that both $h_1$ and $h_3$ have a neighbor in $P \setminus
\{p_1\}$.  If they both have a neighbor in $P \setminus \{p_1, p_2\}$,
then there is a shortest subpath $P'$ of $P \setminus \{p_1, p_2\}$
whose one endnode is adjacent to $h_1$ and the other to $h_3$, and
hence $P' \cup \{p_1, x, h_1, h_3 \}$ induces a square-$3PC(h_1,
h_3)$.  So we may assume w.l.o.g. that $p_2$ is the unique neighbor of
$h_1$ in $P \setminus \{p_1\}$.  Let $p_t$ be the node of $P$ with
lowest index adjacent to $h_3$.  If $t=2$, then Claim~\ref{cl:typeu}
implies $p_2 \in H_1$, a contradiction.  So $t>2$, and hence $\{p_1,
\ldots, p_t, h_1, h_3, x \}$ induces a $3PC(p_1p_2h_1, h_3)$.
Therefore, not both $h_1$ and $h_3$ can have a neighbor in $P\setminus
\{p_1\}$.

Assume w.l.o.g. that $h_1$ has a neighbor in $P \setminus \{p_1\}$.
Suppose that $p_2$ is the unique neighbor of $h_1$ in $P \setminus
\{p_1\}$.  If a node $h_t$, $4<t<n$, has a neighbor in $P$, then $P
\cup \{h_1, h_3, h_t, x \}$ contains a $3PC(h_1p_1p_2, x)$.  So no
node of $h_5\cdots h_{n-1}$ has a neighbor in $P$.  If $h_n$ has no
neighbor in $p_1\cdots p_j$, then $i=4$ and hence $\{p_1, \ldots,
p_j, x \} \cup (H \setminus \{h_2, h_3\})$ induces an odd wheel with
center $x$.  So $h_n$ has a neighbor in $p_1\cdots p_j$.  Let $p_t$ be
such a neighbor with smallest index.  If $t<j$ or $t=j$ and $p_j$ is
not adjacent to $h_4$, then $\{p_1, \ldots, p_t, x\} \cup (H
\setminus \{h_1, h_2 \})$ induces an odd wheel with center $x$.  So
$t=j$ and $p_j$ is adjacent to $h_4$.  Hence $\{p_1, \ldots, p_j,
h_3, h_4, h_n, x \}$ induces a $3PC(h_3h_4x, p_j)$.  Therefore, $h_1$
has a neighbor in $P \setminus \{p_1, p_2 \}$.

Now suppose that $h_1$ does not have a neighbor in $Q \setminus
\{q_1\}$.  Then $h_3$ must have a neighbor in $Q \setminus \{q_1,
q_2\}$, and hence $(P \setminus \{p_1, p_2\}) \cup (Q \setminus \{q_1,
q_2 \}) \cup \{h_1, h_3 \}$ contains a chordless path $R$ from $h_1$
to $h_3$.   
If $p_1$ does not have a neighbor in $Q \setminus \{ q_1,q_2 \}$,
then $R \cup \{p_1, x\}$ induces a square-$3PC(h_1,
h_3)$.  
So $p_1$ has a neighbor in $Q \setminus \{ q_1,q_2\}$. Let $Q'$ be the
shortest path from $p_1$ to $u$ whose vertices are contained in 
$(Q \setminus \{ q_1,q_2\}) \cup \{ p_1\}$. Now apply the same argument to
$P$ and $Q'$.
This proves the claim.  \qed

 
Suppose that $H_1 \neq \emptyset$.  Let $u$ be any node of $C$ that
has a neighbor in $h_4 \cdots h_n$.  Then there exists a path
$P=p_1\cdots p_k$ in $C$ such that $p_k=u$, $p_1 \in H_1$, and no node
of $P \setminus \{p_1\}$ belongs to $H_1$.  By Claim~\ref{upk},
exactly one of $h_1, h_3$ has a neighbor in $P \setminus \{p_1\}$.  If
$h_1$ does then we say that $u$ is \emph{labeled 1} w.r.t.~$H_1$, and
otherwise $u$ is \emph{labeled 3} w.r.t.~$H_1$.  Note that by
Claim~\ref{upk} this label is unique.
\begin{claim}\label{h4hn}
If $H_1\neq\emptyset$, every node of $C$ adjacent to $h_4$ must be
labeled 3 w.r.t.~$H_1$, and every node of $C$ adjacent to $h_n$ must
be labeled 1 w.r.t.~$H_1$.  
\end{claim}
{\em Proof.} Suppose, up to symmetry, that some node $u$ of $C$
adjacent to $h_4$ is labeled 1 w.r.t.~$H_1$.  Let $P=p_1\cdots p_k$ be
a path of $C$ such that $p_1 \in H_1$, $p_k=u$, and no node of $P
\setminus \{p_1\}$ belongs to $H_1$.  Then $h_3$ has no neighbor in $P
\setminus \{p_1\}$.  Let $p_j$ be the node of $P$ with lowest index
adjacent to $h_4$.  By Claim~\ref{upk}, we have $j\ge 3$ and $h_1$ has
a neighbor in $p_3 \cdots p_j$.  But then $\{p_1, p_3, \ldots, p_j,
h_1, h_3, h_4, x \}$ contains a $3PC(h_3h_4x, h_1)$.  This proves the
claim.  \qed

\begin{claim}\label{cl:3}
No node of $C$ is of Type~3.
\end{claim}
{\em Proof.} Assume the contrary.  W.l.o.g. $H_1 \neq \emptyset$.

Suppose that $H_3=\emptyset$.  Let $P=p_1\cdots p_k$ be a shortest
path in $C$ from a node $p_1$ of $H_1$ to a node $p_k$ adjacent to
$h_4$.  So no node of $P \setminus \{p_1\}$ belongs to $H_1$, and no
node of $P \setminus \{p_k\}$ is adjacent to $h_4$.  By
Claims~\ref{upk} and~\ref{h4hn}, we have $k\ge 3$, node $h_3$ has a
neighbor in $P \setminus \{p_1, p_2 \}$, node $h_1$ has no neighbor in
$P \setminus \{p_1\}$, and so, by Claim~\ref{h4hn} again, $h_n$ has no
neighbor in $P$.  Then $h_5$ has no neighbor in $P$, else by
Claim~\ref{pyz}, applied to $h_4$ and $P$, we should have $p_k\in
H_3$.  Some node of $h_6, \ldots, h_{n-1}$ must have a neighbor in
$P$, else $P \cup (H \setminus \{h_2, h_3\}) \cup \{x\}$ induces an
odd wheel with center $x$.  Let $h_i$ be such a node with smallest
index.  If $i$ is odd, then $P \cup \{h_4, \ldots, h_i, x \}$ contains
an odd wheel with center $x$.  So $i$ is even.  If $h_i$ has a
neighbor in $P \setminus \{p_k\}$, then $(P \setminus \{p_k\}) \cup
\{h_3, \ldots, h_i, x\}$ contains an odd wheel with center $x$.  So
$p_k$ is the only neighbor of $h_i$ in $P$.  By Claims~\ref{cl:typeu}
and~\ref{cl:nopyr}, node $h_3$ cannot be adjacent to $p_k$.  But then
$P \cup \{h_3, h_4, h_i, x \}$ contains a 3PC$(xh_3h_4, p_k)$, a
contradiction.  So $H_3\neq \emptyset$.  

Repeating this argument, we obtain that $H_i\neq \emptyset$ for each
odd $i$.

Let $y \cdots z$ be any shortest path in $C$ from a node $y$ of $H_1$
to a node $z$ of $H_3$.  By Claim~\ref{h4hn}, $z$ is labeled 3
w.r.t.~$H_1$, and $y$ is labeled 3 w.r.t.~$H_3$.  So there exists a
largest odd integer $i$ such that $C$ contains a chordless path $P=
p_1 \cdots p_k$ from a node $p_1$ of $H_1$ to a node $p_k$ of $H_i$,
with no intermediate nodes in $H_1 \cup H_i$, such that $p_k$ is
labeled 3 w.r.t.~$H_1$ and $p_1$ is labeled $i$ w.r.t.~$H_i$.  In
particular, by Claim~\ref{upk}, we have $k\ge 3$, node $h_i$ has a
neighbor in $P \setminus \{p_k, p_{k-1}\}$ and $h_{i+2}$ has no
neighbor in $P \setminus \{p_k\}$.  Also since $p_k$ is labeled 3
w.r.t.~$H_1$, and since by Claim~\ref{h4hn} all nodes of $H_{n-1}$ are
labeled 1 w.r.t.~$H_1$, it follows that $i<n-1$.  Since $H_{i+2}\neq
\emptyset$, there exists a shortest path $Q=q_1\cdots q_l$ in $C$ such
that $q_1\in H_{i+2}$ and $q_l$ has a neighbor in $P$.  Node $q_1$ is
not adjacent to $p_k$, for otherwise $\{x, h_{i+1}, h_{i+2}, h_{i+3},
p_k, q_1\}$ induces an odd wheel.

Suppose that $q_l$ is not adjacent to $p_k$.  If $q_l$ has a neighbor
in $P \setminus \{p_k, p_{k-1}\}$, then $(P \setminus \{p_{k-1}\})
\cup Q \cup \{h_i, h_{i+2}, x \}$ contains a square-$3PC(h_i,
h_{i+2})$.  So $N(q_l) \cap P=\{p_{k-1}\}$.  Let $Q'=q_1'\cdots q_t'$
be the path induced by $(P \setminus \{p_k\}) \cup Q$, where $q_1'=
q_1$ and $q_t'=p_1$.  Suppose that $q_1$ is labeled 1 w.r.t.~$H_1$.
By Claim~\ref{upk} applied to $Q'$, this is possible only if $p_k$ is
the only neighbor of $h_3$ in $P$, so $i=3$, and $h_1$ has a neighbor
in $Q$.  Then $k=3$, for otherwise $Q\cup \{x, h_1, h_3, p_1, p_{k-1},
p_k\}$ contains a square-$3PC(h_1, h_3)$.  But then $\{x, h_3, h_6\}
\cup P\cup Q$ contains a square-$3PC(h_3, p_2)$.  So $q_1$ is labeled
3 w.r.t.~$H_1$.  If $p_1$ is labeled $i+2$ w.r.t.~$H_{i+2}$, the
maximality of $i$ is contradicted.  Hence $p_1$ is labeled $i+4$
w.r.t.~$H_{i+2}$, and by Claim~\ref{upk}, we have $t\ge 3$, node
$h_{i+4}$ has a neighbor in $Q' \setminus \{q_1', q_2'\}$, and
$h_{i+2}$ has no neighbor in $Q' \setminus \{q_1'\}$.  If $l>1$, then
$(Q' \setminus \{q_2'\}) \cup \{p_k, h_{i+2}, h_{i+4}, x \}$ contains
a square-$3PC(h_{i+2}, h_{i+4})$.  So $l=1$.  But then $P \cup \{q_1,
h_2, h_{i+2}, x \}$ contains a square-$3PC(p_{k-1}, h_{i+2})$.
Therefore, $q_l$ must be adjacent to $p_k$.  Thus $l\ge 2$.

Let $p_j$ be the node of $P$ with smallest index adjacent to $q_l$.
Let $Q'=q_1'\cdots q_t'$ be the path induced by $Q \cup \{p_1, \ldots,
p_j\}$, where $q_1'=q_1$ and $q_t'=p_1$.  Note that $h_1$ cannot have
a neighbor in $Q$, since otherwise $(P \setminus \{p_2\}) \cup Q \cup
\{h_1, h_3, x \}$ contains a square-$3PC(h_1, h_3)$.  So node $q_1'$
must be labeled 3 w.r.t.~$H_1$.  Node $q_t'$ must be labeled $i+4$
w.r.t.~$H_{i+2}$, else the maximality of $i$ is contradicted.  So, by
Claim~\ref{upk}, we have $t\ge 3$, node $h_{i+4}$ has a neighbor in
$Q' \setminus \{q_1', q_2'\}$, and $h_{i+2}$ has no neighbor in
$Q'\setminus \{q_1'\}$.  But then $P \cup (Q \setminus \{q_2\}) \cup
\{h_{i+2}, h_{i+4}, x \}$ contains a square-$3PC(h_{i+2}, h_{i+4})$.
This proves the claim.  \qed

By Claims~\ref{cl:typeu}, \ref{cl:nopyr} and \ref{cl:3}, the nodes of
$C$ that have a neighbor in $H$ are of Type~1, 2 or 5.  Let $C'$ be a
minimal connected induced subgraph of $C$ such that for some $s, t\in
\{1, \ldots, n\}$ node $h_t$ is not adjacent to $h_s$ nor $h_{s+1}$,
and each of $h_s, h_{s+1}, h_t$ has a neighbor in $C'$.  W.l.o.g.
$s=1$.  Let $P=p_1\cdots p_k$ be a shortest path in $C'$ such that
$h_1$ is adjacent to $p_1$ and $h_2$ to $p_k$.   
 
Suppose that $k=1$.  So $p_1$ is of Type~2.  Let $Q=q_1 \cdots q_l$ be
a path in $C'$ such that $q_1$ has a neighbor in $H\setminus \{h_n,
h_1, h_2, h_3\}$ and $q_l$ is adjacent to $p_1$.  Thus $C'= Q\cup
\{p_1\}$, and no node of $Q\setminus \{q_1\}$ has a neighbor in $H
\setminus \{h_n, h_1, h_2, h_3\}$.  If both $h_1, h_2$ have a neighbor
in $Q$, then $Q$ contradicts the minimality of $C'$.  So we may assume
that $h_1$ has no neighbor in $Q$.  Then $h_n$ has no neighbor in
$Q\cup \{p_1\}$, for otherwise a subpath of $Q\cup \{p_1\}$ violates
Claim~\ref{cl:typep} or~\ref{cl:nopyr}.  If $h_2$ too has no neighbor
in $Q$, then similarly $h_3$ has no neighbor in $Q\cup \{p_1\}$, and
then by Claim~\ref{cl:typep}, $H \cup Q \cup p_1$ induces a
$3PC(\Delta, \Delta )$, contradicting Claim~\ref{cl:nopyr}.  So $h_2$
has a neighbor in $Q$.  Let $h_t$ be the node of $H \setminus \{h_n,
h_1, h_2, h_3\}$ with highest index adjacent to $q_1$.  Then $Q \cup
\{h_t, \ldots, h_n, h_1, h_2\}$ and $Q \cup \{h_1, h_2, h_t, x\}$
induce two wheels with center $h_2$, one of which is odd, a
contradiction.  So $k>1$.

Let $Q=q_0\cdots q_l$ be a shortest path such that $q_0\in H\setminus
\{h_n, h_1, h_2, h_3\}$, $Q\setminus \{q_0\}\subseteq C'$, and $q_l$
has a neighbor in $P$ (possibly $l=0$).  So if $l>0$, no node of
$P\cup Q\setminus \{q_0, q_1\}$ has a neighbor in $H\setminus \{h_n,
h_1, h_2, h_3\}$.  Note that $C'=P\cup Q\setminus \{q_0\}$.

Suppose that $h_1$ has a neighbor in $Q$.  So $l>0$.  Let $h_t$ be the
node of $h_4\cdots h_{n-1}$ with smallest index adjacent to $q_1$.
Then, by the minimality of $C'$, $N(q_l) \cap P=\{ p_1\}$ and $h_2$ has no
neighbor in $Q$.  If $h_3$ has no neighbor in $P\cup Q$, then $P\cup
(Q\setminus \{ q_0\})
\cup \{h_1, h_2, \ldots, h_t\}$ 
or $P \cup (Q \setminus \{ q_0 \}) \cup \{ x,h_1,h_2,h_t \}$
induces an odd wheel with center
$h_1$.  So $h_3$ has a neighbor in $P\cup Q$.  Then Claim~\ref{pyz},
applied to $h_2$ and $P\cup Q$, implies that $p_k$ is of Type 3, a
contradiction.  Therefore $h_1$ cannot have a neighbor in $Q$, and
similarly neither can $h_2$.

If $q_l$ has exactly one neighbor $p_a$ in $P$, then $P \cup Q \cup
\{h_1, h_2, x\}$ induces a $3PC(h_1h_2x, p_a)$.  If $q_l$ has two
non-consecutive neighbors in $P$, then the same node set contains a
$3PC(h_1h_2x, q_l)$.  So $N(q_l) \cap P=\{p_a, p_{a+1}\}$.  Node $h_3$
must have a neighbor in $P$, else $P \cup Q \cup \{h_1, h_2, \ldots,
q_0\}$ contains a $3PC(q_lp_ap_{a+1}, h_2)$.  By Claim~\ref{cl:typeu},
$h_3$ cannot be adjacent to $p_1$.  Now $h_2, h_3, q_0$ all have a
neighbor in $C' \setminus \{p_1\}$, which is connected, and so the
minimality of $C'$ implies $q_0=h_4$.  Then $N(h_3) \cap P=\{ p_k\}$, else
$h_1, h_3, h_4$ all have a neighbor in $C' \setminus \{p_k\}$,
contradicting the minimality of $C'$.  But then $P \cup \{h_1, h_2,
h_3, x\}$ induces an odd wheel with center $h_2$.  This completes the
proof of the theorem.  \qed


\section{On the complexity of several detection problems}
\label{sec:reco}

\subsection{Detecting a $3PC(\cdot,\cdot)$ or a $3PC(\Delta, \cdot)$}

Chudnovsky and Seymour gave an $O(n^{11})$ to decide if a graph
contains a $3PC(\cdot, \cdot)$ \cite{chudnovsky.seymour:theta} and an
$O(n^9)$ algorithm to decide if a graph contains a $3PC(\Delta,
\cdot)$ \cite{chudnovsky.c.l.s.v:reco}.  Here we give an $O(n^7)$
algorithm that decides whether a given graph has either a $3PC(\cdot,
\cdot)$ or a $3PC(\Delta, \cdot)$.  Say that a $3PC(\Delta, \cdot)$
is~\emph{long} if its three paths have length at least~2.  Otherwise,
exactly one of its paths has length~1, and we say it is \emph{short}.
Here is a sufficient condition for a graph to have a $3PC(\cdot,
\cdot)$ or a long $3PC(\Delta, \cdot)$.

\begin{lemma}
  \label{lemma:lb}
Let $G$ be a graph with four nodes $u, a, b, c$ and a set
$W\subseteq V(G)\setminus\{u, a, b, c\}$ such that:
  \begin{itemize}
    \item
      $\{u, a, b, c\}$ induces a claw centered at $u$;
    \item
      $W$ induces a connected subgraph of $G$;
    \item 
      $u$ has no neighbour in $W$;
    \item
      Every node in $\{a,b,c\}$ has exactely one neighbour in $W$.
  \end{itemize}
Then, $G$ contains a $3PC(\cdot, \cdot)$ or a long $3PC(\Delta,
\cdot)$.
\end{lemma}
\emph{Proof.}  
Let $a', b', c'$ be the neighbours of $a,b,c$ in $W$.  If $a'=b'=c'$,
then $\{ u,a,b,c,a' \}$ induces a $3PC(u,a')$.  Now we may assume $a'
\neq b'$.  Then $G[W]$ contains a path with ends $a'$ and $b'$, and we
let $P$ be a shortest such path.  Let $Q = c', \ldots ,v$ be a path in
$G[W]$ such that $v$ has a neighbour in $P$ and no node of $Q\setminus  v$
has a neighbor in $P$.  Such a $Q$ exists because $W$ is connected
(possibly $c'=v$).  Let $d$ (respectively $e$) be the neighbour of $v$
in $P$ closest to $a'$ (respectively to $b'$).  Let $T$ be the graph
induced by $P\cup Q \cup \{ u,a,b,c \}$.  If $d=e$, then $T$ is a
$3PC(d,u)$.  If $d, e$ are distinct and adjacent then $T$ is a long
$3PC(vde,u)$.  If $d, e$ are distinct and non-adjacent then $T$
contains a $3PC(v,u)$.
\qed 

Now we can give an $O(n^6)$ algorithm for the detection of
non-square-$3PC(\cdot, \cdot)$'s or long $3PC(\Delta, \cdot)$'s, very
similar to an $O(n^5)$ algorithm by Maffray and
Trotignon~\cite{maffray.t:reco} that detects $3PC(\Delta, \Delta)$'s or
$3PC(\Delta, \cdot)$'s:

\label{algo:long}

\begin{description}
\item[\sc Input:] A graph $G$.

\item[\sc Output:] 
The positive answer ``$G$ contains a non-square $3PC(\cdot, \cdot)$ or
a long $3PC(\Delta, \cdot)$'' if it does; else the negative answer
``$G$ contains no non-square-$3PC(\cdot, \cdot)$ and no long
$3PC(\Delta, \cdot)$.''

\item[\sc Method:] 
For every claw $\{u, b_1, b_2, b_3\}$ centered at $u$ do:

Step 1.  Compute the set $X_1$ of those nodes of $V(G)$ that are
adjacent to $b_1$ and not adjacent to $u$, $b_2$ or $b_3$, and the
similar sets $X_2, X_3$, and compute the set $X$ of those nodes of
$V(G)$ that are not adjacent to any of $u, b_1, b_2, b_3$.  Compute
the connected components of $X$ in $G$.  For each component $H$ of
$X$, and for $i=1, 2, 3$, if some node of $H$ has a neighbour in $X_i$
then mark $H$ with label $i$.

Step 2.  For every component $H$ of $X$ that has received label $i\in
\{1, 2, 3\}$, and for every node $x$ of $X_i$ that has a neighbour in
$H$, assign to $x$ the other labels of $H$ (if any).  For each $i=1,
2, 3$ and for every node $x$ of $X_i$ that has a neighbour in $X_j$
with $j\in \{1, 2, 3\}$ and $j\neq i$, assign label $j$ to $x$.

Step 3.  If some node of $X_1\cup X_2\cup X_3$ gets two different
labels, return the positive answer and stop.

If the positive answer has not been returned at step 3, return the
negative answer.
    
\item[\sc Complexity:] $O(n^6)$.
\end{description}
\emph{Proof of correctness.} Suppose that $G$ contains a 
non-square-$3PC(\cdot, \cdot)$ 
or a long $3PC(\Delta, \cdot)$, say $K$.  Let $u,
b_1, b_2, b_3$ be the four nodes of a $u$-centered claw of $K$, and
for $i=1, 2, 3$ let $c_i$ be the neighbour of $b_i$ in $K \setminus
\{u, b_1, b_2, b_3\}$.  Let us observe what the algorithm will do when
it examines the 4-tuple $\{u, b_1, b_2, b_3\}$.  The algorithm will
place the three nodes $c_1, c_2, c_3$ in the sets $X_1, X_2, X_3$
respectively.

First suppose that $K$ is neither a $3PC(\cdot , \cdot )$ with one of
the paths of length 2 nor a $3PC(\Delta , \cdot )$ with all three
paths of length 2.  Then $K \setminus \{ u,b_1,b_2,b_3,c_1,c_2,c_3 \}$
is contained in a connected component $H$ of $X$, and all three nodes
$c_1,c_2,c_3$ have a neighbor in $H$, i.e. $H$ gets assigned all three
labels 1, 2 and 3.  So $c_1$ gets labels 2 and 3, and hence step~3
returns the positive answer.  If $K$ is a $3PC(\Delta , \cdot )$ with
all three paths of length 2, then $K$ is a $3PC(c_1c_2c_3,u)$, so by
step~2 $c_1$ gets labels 2 and 3, and hence step~3 returns the
positive answer.  Finally assume that $K$ is a $3PC(\cdot , \cdot )$
with one of the paths of length 2.  W.l.o.g. $K$ is a $3PC(u,c_1)$.
Let $H_2$ (resp.~$H_3$) be the connected component of $K \setminus \{
u,b_1,b_2,b_3,c_1,c_2,c_3 \}$ in which both $c_1$ and $c_2$
(resp.~$c_1$ and $c_3$) have a neighbor.  Since $c_1$ has a neighbor
in both $H_2$ and $H_3$, in step~2 $c_1$ gets labels 2 and 3, and
hence step~3 returns the positive answer.

Conversely, suppose that the algorithm returns the positive answer
when it is examining a $u$-centered claw $\{x, b_1, b_2, b_3\}$.  So
(up to symmetry) some node $c_1\in X_1$ gets labels $2$ and $3$ at
step~2.  This means that for $j=2, 3$, there exists a path $R_j$ from
$c_1$ to a node of $X_j$ such that the interior nodes of $R_j$ (if
any) lie in $X$.  We can apply Lemma~\ref{lemma:lb} to the to the claw
$\{u, b_1, b_2, b_3\}$ and the set $W=V(R_2)\cup V(R_3)$, which
implies that this subgraph (and thus $G$ itself) contains a long
$3PC(\Delta, \cdot)$ or a $3PC( \cdot , \cdot)$ (that is non-square
because the $c_i$'s are pairwise distinct).  This completes the proof
of correctness.

Finding all 4-tuples takes time $O(n^4)$.  For each 4-tuple, computing
the sets $X_1,$ $X_2,$ $X_3,$ $X$ takes time $O(n^2)$.  Finding the
components of $X$ takes time $O(n^2)$.  Marking the components at the
end of step 1 can be done as follows: for each edge $uv$ of $G$, if
$u$ is in a component $H$ of $X$ and $v$ is in some $X_i$ then mark
$H$ with label $i$.  This takes time $O(n^2)$.  Marking the nodes of
$X_1\cup X_2\cup X_3$ at step 2 can be done similarly.  Thus the
overall complexity is $O(n^6)$.  \qed

Detecting square-$3PC(\cdot, \cdot)$'s is easy to do in time $O(n^6)$
as noted in the introduction.  So, in order to detect $3PC(\cdot,
\cdot)$'s or $3PC(\Delta, \cdot)$'s, we are left with the problem of
deciding whether a graph has  a short $3PC(\Delta, \cdot)$.
This is an NP-complete problem (proved in the next section) but we can
solve it assuming that the graph has no $3PC(\cdot, \cdot)$ and no
long $3PC(\Delta, \cdot)$.  This could be done in time $O(n^9)$ using
the algorithm of Chudnovsky and Seymour
\cite{chudnovsky.c.l.s.v:reco} that detects $3PC(\Delta, \cdot)$.  We
propose here something faster and simpler but based on the same ideas.

If $K$ is a short $3PC(bde, u)$ such that $u$ sees $b$, $a$ is the
neighbour of $u$ along the path of $K$ from $u$ to $d$, and $c$ is the
neighbour of $u$ along the path of $K$ from $u$ to $e$, then we say
that $(u,a,b,c,d,e)$ is a \emph{frame} of $K$.

\begin{lemma}
\label{lemma:rep}
Let $G$ be a graph with no $3PC(\cdot, \cdot)$ and no long
$3PC(\Delta, \cdot)$.  Let $K$ be a smallest short $3PC(\Delta,
\cdot)$ of $G$ with frame $(u,a,b,c,d,e)$.  Let $P$ be the path of
$K\setminus\{u\}$ between $a$ and $d$.  Let $R$ be a shortest path of
$G$ between $a$ and $d$ such that the interior nodes of $R$ are not
adjacent to $b,c,e$.  Then the graph induced by $(K \setminus P) \cup
R$ induces a smallest short $3PC(\Delta, \cdot)$ of $G$.
\end{lemma}
\emph{Proof.} Note that $G$ has no $3PC(\Delta, \cdot)$ smaller that
$K$, because it has no long $3PC(\Delta, \cdot)$ at all, and because
$K$ is a smallest short $3PC(\Delta, \cdot)$.  Let us denote by $r_1 =
a, \ldots, r_k = d$ the nodes of $R$.  Let $r_s$ the neighbour of $u$
in $R$ with greatest index.  Note that $r_s$ exists because $r_1$ is a
neighbor of~$u$.

We claim that the graph induced by $(K \setminus P) \cup \{ r_s,
\ldots, r_k\}$ induces a short $3PC(\Delta, \cdot)$.  Indeed, let $Q$
be the path of $K\setminus\{u\}$ with end-nodes $c$ and $e$.  Let us
denote by $q_1=c, \ldots, q_l = e$ the nodes of $Q$.  If no node of
$r_s \ldots r_{k-1}$ has neighbours in $Q$, then the claim holds.  So,
we may assume that there is a node $r_t$ in $ r_s \ldots r_{k-1}$ that
has a neighbour in the interior of $Q$, and we choose $t$ maximum.
Note that $t>1$.  Let $i$ be the smallest index and $j$ be the
greatest index such that $q_i, q_j$ are neighbors of $r_t$.  If $t =
s$, then $\{r_t, \ldots, r_k, q_j, \ldots, q_l, u, b\}$ induces a
$3PC(bde, r_t)$ that is smaller than $K$, a contradiction.  So $t> s$.
If $i = j$ then $Q \cup \{r_t, \ldots, r_k, u, b\}$ induces a $3PC(bde,
q_j)$ that is smaller than $K$, a contradiction.  If $j > i+1$ then
$\{r_t, \ldots, r_k, q_1, \ldots, q_i, q_j, \ldots, q_l, u, b\}$ induces a
$3PC(bde, r_t)$ that is smaller than $K$, a contradiction.  So $j =
i+1$.  There is a shortest path $S$ with end-nodes $r_t$ and $a$ in
the graph induced by $P \cup \{r_t \ldots, r_k\}$.  If $d\notin V(S)$,
then $Q \cup S \cup \{u, b\}$ induces a long $3PC(r_t q_i q_j, u)$, a
contradiction.  If $d \in V(S)$ then $S \cup \{q_1, \ldots, q_i, b,
u\}$ induces a $3PC(d, u)$, a contradiction.  This proves the claim.

We proved that $K' = (K \setminus P) \cup \{ r_s \ldots r_k \}$ induces
a $3PC(\Delta, \cdot)$.  If $s > 1$, then $K'$ is a $3PC(\Delta,
\cdot)$ smaller than $K$, a contradiction.  So, $s = 1$ proving the
lemma.  \qed

Now we can give an algorithm for the detection of short $3PC(\Delta,
\cdot)$'s in graphs with no $3PC(\cdot, \cdot)$ and no long
$3PC(\Delta, \cdot)$:

  \label{algo:short}
  \begin{description}
    
  \item{\sc Input:}  A graph  $G$ with no  $3PC(\cdot, \cdot)$  and no
    long $3PC(\Delta, \cdot)$.

  \item{\sc  Output:}  The  positive  answer ``$G$  contains  a  short
    $3PC(\Delta, \cdot)$'' if it  does; else the negative answer ``$G$
    contains no short $3PC(\Delta, \cdot)$.''

  \item{\sc  Method:}   

    For every 5-tuple $(a, b, c, d, e)$ of nodes do:

    Step 1.  Compute in $V(G)  \setminus (N(b) \cup N(c) \cup N(e))$ a
    shortest  path $P$ from  $a$ to  $d$ (if  any).  Compute  in $V(G)
    \setminus (N(b) \cup N(a) \cup N(d))$ a shortest path $Q$ from $c$
    to $e$ (if any).  If at least one of the paths  does not exist, go
    to the next 5-tuple.
    
    Step 2. Check if the edge-set of $G[V(P) \cup V(Q) \cup \{a, b, c,
    d, e\}]$  is exactly $E(P) \cup  E(Q) \cup \{bd, be,  de\}$. If it
    does not, go to the next 5-tuple.

    Step 3. For every node $u$  of $G$, check if $ua$, $ub$ and $uc$
    are the only edges  from $u$ to $V(P) \cup V(Q) \cup  \{a, b, c, d,
    e\}$.  If  it  is  the   case,  return  the  positive  answer  and
    stop. Else, go to the next 5-tuple.
   
    If after checking every 5-tuple,  the positive answer has not been
    returned, return the negative answer.

  \item{\sc Complexity:} $O(n^7)$.

  \end{description}
\emph{Proof.} If the algorithm gives the positive answer, let us
consider the 5-tuple $(a, b, c, d, e)$, the paths $P, Q$, and the node
$u$ that make the algorithm stop.  It is clear by the method that
$V(P) \cup V(Q) \cup \{a, b, c, d, e\}$ induces a short $3PC(\Delta,
\cdot)$.  Conversely, if $G$ has a short $3PC(\Delta, \cdot)$ $K$ with
frame $(u, a, b, c, d, e)$, then let us examine what the algorithm
will do when checking the 5-tuple $(a, b, c, d, e)$.  By two
applications of lemma~\ref{lemma:rep}, we see that the two paths
computed by the algorithm can take the place of the corresponding
paths of $K$, to give another short $3PC(\Delta, \cdot)$ $K'$
(possibly not $K$) with apex $u$.  So, since $u$ exists, the algotithm
will find a node that has same neighbourhood that $u$ (in $K'$) and
give the positive answer.  This proves the correctness of the
algorithm.  Checking every 5-uple takes $O(n^5)$, compute shortest
paths takes $O(n^2)$, checking every possible edge at step 2 takes
$O(n)^2$, checking every $u$ at step 3 take $O(n)$, and checking every
neighbour of $u$ takes $O(n)$.  So the overall complexity is $O(n^7)$.
\qed

By the algorithms above, we obtain:

\begin{theorem}
  There is an $O(n^7)$-time algorithm that decides whether a
  graph has a $3PC(\cdot, \cdot)$ or a $3PC(\Delta, \cdot)$.
\end{theorem}

\subsection{NP-completeness results}

Let us call problem $\Pi$ the decision problem whose input is a graph
$G$ and two non-adjacent nodes $a,b$ of $G$ of degree $2$ and whose
question is: ``Does $G$ have a hole that contains both $a,b$?''
Bienstock~\cite{bienstock:evenpair} proved that this problem is
NP-complete.  Adapting Bienstock's proof, Maffray and
Trotignon~\cite{maffray.t:reco} remarked that the problem remains
NP-complete for triangle-free graphs.  Here is an easy consequence:

\begin{theorem}
  \label{npcwheel}
  The problem of deciding whether a graph has an odd wheel is
  NP-complete. The problem of deciding whether a graph has a short
  $3PC(\Delta , \cdot )$ is NP-complete.
\end{theorem}

\emph{Proof.} Suppose there is a polynomial time algorithm $\cal A$
for the detection of short $3PC(\Delta, \cdot)$'s or an algorithm
$\cal A'$ for the detection of odd wheels.  Let $G, a, b$ be an
instance of $\Pi$.  Let $b', b''$ be the neighbours of $b$ in $G$.
Build a graph $H$ by adding to $G$ nodes $c_1, c_2, c_3, c_4$, $c_5$
and edges $c_1a,$ $c_1c_2,$ $c_1c_3,$ $c_2c_3,$ $c_2c_4,$ $c_4b',$
$c_3c_5,$ $c_5b''$.  Since $G$ has no triangle, every short
$3PC(\Delta, \cdot)$ in $H$ has apex $a$.  So there is a short
$3PC(\Delta, \cdot)$ in $H$ if and only if there is a hole passing
through $a$ and $b$ in $G$.  Similarly, there is an odd wheel in $H$
if and only if there is a hole passing through $a$ and $b$ in $G$.
Thus, Algorithm $\cal A$ (or $\cal A'$) yields a polynomial time
algorithm that solves the NP-complete problem~$\Pi$.  \qed

\

When $k\geq 2$, a $kPC(\Delta, \cdot)$ is a graph induced by $k$
chordless paths $P_1, \ldots, P_k$, such that each path $P_i$ is from
a node $x$ to a node $y_i \neq x$, the nodes $y_1,\ldots, y_k$ are
distinct and pairwise adjacent, and the union of any two paths $P_i,
P_j$ induces a hole.  Note that this latter condition implies that at
most one of the paths $P_1, \ldots, P_k$ can have length $1$.

When $k\geq 2$, a $kPC(\cdot, \cdot)$ is a graph induced by $k$
chordless paths $P_1, \ldots, P_k$, such that they all have the same endnodes,
and the union of any two paths $P_i,
P_j$ induces a hole.

For any integer $n\ge 1$, let $K_{1, n}$ denote the graph on $n+1$
nodes with $n$ edges and a node of degree $n$.  Adapting the proof of
Bienstock, we prove that $\Pi$ remains NP-complete for $K_{1, 4}$-free
graphs.  Before presenting the proof of this result, we point out that
Problem $\Pi$ is polynomial for $K_{1, 3}$-free graphs.  To see this,
consider an instance $(G, a, b)$ of $\Pi$ where $G$ is $K_{1,
  3}$-free.  Consider the graph $G'$ obtained from $G$ by adding a
node $c$ of degree~$2$ adjacent to $a$ and $b$.  It is easy to see
that $G$ contains a hole going through $a$ and $b$ if and only if $G'$
contains a $3PC(\cdot, \cdot)$.  Since this last problem is polynomial
\cite{chudnovsky.seymour:theta}, $\Pi$ is polynomial when restricted
to $K_{1, 3}$-free graphs.
\begin{theorem}
\label{thm:pinpc}
Problem $\Pi$ is NP-complete for $K_{1, 4}$-free graphs.
\end{theorem}
\emph{Proof.} Let us give a polynomial reduction from the problem {\sc
$3$-Satisfiability} of Boolean functions to problem $\Pi$ restricted
to $K_{1, 4}$-free graphs.  Recall that a Boolean function with
$n$ variables is a mapping $f$ from $\{0, 1\}^n$ to $\{0, 1\}$.  A
Boolean vector $\xi\in\{0, 1\}^n$ is a \emph{truth assignment} for $f$
if $f(\xi)=1$.  For any Boolean variable $z$ on $\{0, 1\}$, we write
$\overline{z}:=1-z$, and each of $z, \overline{z}$ is called a
\emph{literal}.  An instance of {\sc $3$-Satisfiability} is a Boolean
function $f$ given as a product of clauses, each clause being the
Boolean sum $\vee$ of three literals; the question is whether $f$
admits a truth assignment.  The NP-completeness of {\sc
$3$-Satisfiability} is a fundamental result in complexity theory, see
\cite{garey.johnson:np}.

Let $f$ be an instance of {\sc $3$-Satisfiability}, consisting of $m$
clauses $C_1, \ldots, C_m$ on $n$ variables $z_1, \ldots, z_n$.  Let
us build a graph $G_f$ with two specialized nodes $a,b$, such that
there will be a hole containing $a$ and $b$ in $G$ if and only if
there exists a truth assignment for $f$.

\begin{figure}
\begin{center}
\includegraphics{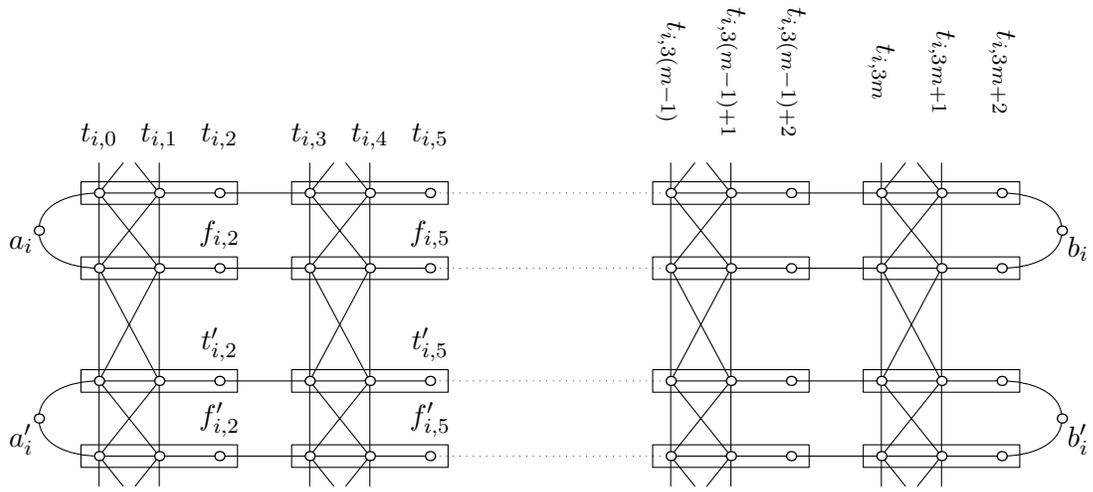}
\end{center}
\caption{Graph $G(z_i)$}\label{fig.reco.1}
\end{figure}

\begin{figure}
\begin{center}
\includegraphics{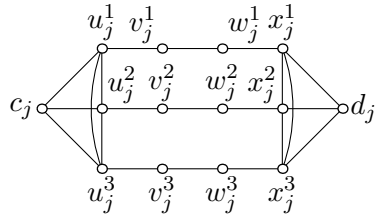}
\end{center}
\caption{Graph $G(C_j)$}\label{fig.reco.2}
\end{figure}

\begin{figure}
\begin{center}
\includegraphics{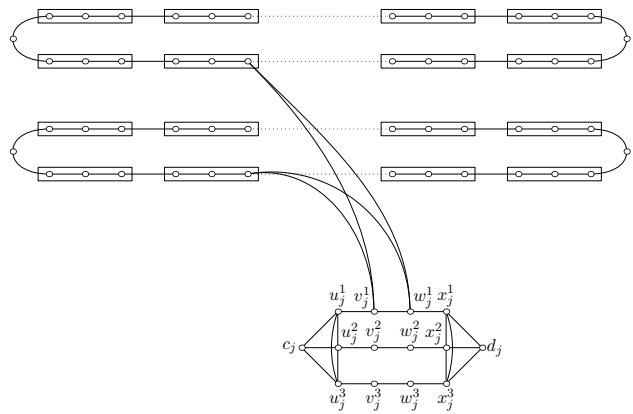}
\end{center}
\caption{The four edges added to $G_f$ in the case 
$y_j^1=z_i$}\label{fig.reco.3}
\end{figure}

\begin{figure}
\begin{center}
\includegraphics{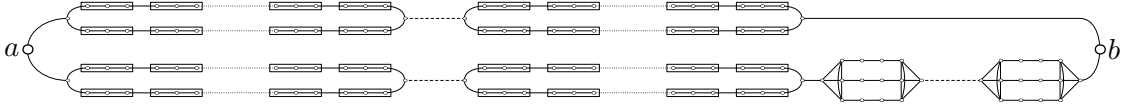}
\end{center}
\caption{Graph $G_f$}\label{fig.reco.4}
\end{figure}

For each variable $z_i$ ($i=1, \ldots, n$), make a graph $G(z_i)$ with
four nodes $a_i,$ $b_i,$ $a'_i,$ $b'_i$, and $4(3(m+1))$ nodes $t_{i,
  j},$ $f_{i, j},$ $t'_{i, j},$ $f'_{i, j}$, with $j \in \{0, \ldots,
3m+2\}$.  Add edges so that the four sets $\{a_i,$ $t_{i, 0},$ $t_{i,
  1}, \ldots,$ $t_{i, 3m+2}, b_i\}$, $\{a_i, f_{i, 0}, f_{i, 1},
\ldots, f_{i, 3m+2}, b_i\}$, $\{a'_i, t'_{i, 0}, t'_{i, 1}, \ldots,
t'_{i, 3m+2}, b'_i\}$, $\{a'_i, f'_{i, 0}, f'_{i, 1}, \ldots, f'_{i,
  3m+2}, b'_i\}$ all induce paths (and the nodes appear in this order
along these paths).  For $k = 0, \dots, m$, add every possible edges
between $\{t_{i, 3k}, t_{i, 3k+1}\}$ and $\{f_{i, 3k}, f_{i, 3k+1}\}$,
between $\{f_{i, 3k}, f_{i, 3k+1}\}$ and $\{t'_{i, 3k}, t'_{i,
  3k+1}\}$, between $\{t'_{i, 3k}, t'_{i, 3k+1}\}$ and $\{f'_{i, 3k},
f'_{i, 3k+1}\}$, between $\{f'_{i, 3k}, f'_{i, 3k+1}\}$ and $\{t_{i,
  3k}, t_{i, 3k+1}\}$.  See Figure~\ref{fig.reco.1}.


For each clause $C_j$ ($j=1, \ldots, m$), with $C_j=y_j^1\vee
y_j^2\vee y_j^3$, where each $y_j^p$ ($p=1, 2, 3$) is a literal from
$\{z_1,$ $\ldots,$ $z_n,$ $\overline{z}_1,$ $\ldots,$
$\overline{z}_n\}$, make a graph $G(C_j)$ with fourteen nodes $c_j,
d_j, u_j^1, v_j^1, w_j^1, x_j^1, u_j^2, v_j^2, w_j^2, x_j^2, u_j^3,
v_j^3, w_j^3, x_j^3$.  Add edges so that the three sets $\{c_j, u_j^1,
v_j^1, w_j^1, x_j^1, d_j\}$, $\{c_j, u_j^2, v_j^2, w_j^2, x_j^2,
d_j\}$, $\{c_j, u_j^3, v_j^3, w_j^3, x_j^3, d_j\}$, all induce paths
(and the nodes appear in this order along these paths).  Add six more
edges so that $\{u_j^1, u_j^2, u_j^3\}$, $\{x_j^1, x_j^2, x_j^3\}$
induce two triangles.
See Figure~\ref{fig.reco.2}.

The graph $G_f$ is obtained from the disjoint union of the $G(z_i)$'s
and the $G(C_j)$'s as follows.  For $i=1, \ldots, n-1$, add edges
$b_ia_{i+1}$ and $b'_ia'_{i+1}$.  Add an edge $b'_nc_1$.  For $j=1,
\ldots, m-1$, add an edge $d_jc_{j+1}$.  Introduce the two special
nodes $a,b$ and add edges $aa_1, aa'_1$ and $bd_m, bb_n$.  See
Figure~\ref{fig.reco.4}.  For $p=1, 2, 3$, if $y_j^p=z_i$, then add
four edges $v_j^pf_{i, 3j-1}, v_j^pf'_{i, 3j-1}$, $w_j^pf_{i, 3j-1},
w_j^pf'_{i, 3j-1}$, while if $y_j^p=\overline{z}_i$ then add four
edges $v_j^pt_{i, 3j-1}, v_j^pt'_{i, 3j-1}$, $w_j^pt_{i, 3j-1},
w_j^pt'_{i, 3j-1}$.  See Figure~\ref{fig.reco.3}.  Clearly the size of
$G_f$ is polynomial (actually quadratic) in the size $n+m$ of $f$.
Moreover, it is a routine matter to check that $G_f$ contains no
$K_{1, 4}$, and that $a,b$ are non-adjacent and both have degree two.

Suppose that $f$ admits a truth assignment $\xi\in\{0, 1\}^n$.  We can
build a hole in $G$ by selecting nodes as follows.  Select $a,b$.  For
$i=1, \ldots, n$, select $a_i, b_i, a'_i, b'_i$.  If $\xi_i=1$ select
$t_{i, j}, t'_{i, j}$ where $j \in \{0, \ldots, 3m+2\}$.  If $\xi_i=0$
select $f_{i, j}, f'_{i, j}$ where $j \in \{0, \ldots, 3m+2\}$.  For
$j=1, \ldots, m$, since $\xi$ is a truth assignment for $f$, at least
one of the three literals of $C_j$ is equal to $1$, say $y_j^p=1$ for
some $p\in\{1, 2, 3\}$.  Then select $c_j, d_j$ and $u_j^p, v_j^p,
w_j^p, x_j^p$.  Now it is a routine matter to check that the selected
nodes induce a cycle $Z$ that contains $a,b$, and that $Z$ is
chordless, so it is a hole.  The main point is that there is no chord
in $Z$ between some subgraph $G(C_j)$ and some subgraph $G(z_i)$, for
that would be either an edge $t_{i, 3j-1}v_j^p$ (or similarly $t'_{i,
  3j-1}v_j^p$, $t_{i, 3j-1}w_j^p$, $t'_{i, 3j-1}w_j^p$) with
$y_j^p=z_i$ and $\xi_i=1$, or, symmetrically, an edge
$f_{i,3j-1}v_j^p$ (or similarly $f'_{i, 3j-1}v_j^p$, $f_{i,
  3j-1}w_j^p$, $f'_{i, 3j-1}w_j^p$) with $y_j^p=\overline{z}_i$ and
$\xi_i=0$, and in either case this would contradict the way the nodes
of $Z$ were selected.

Conversely, suppose that $G_f$ admits a hole $Z$ that contains $a,b$.
Clearly $Z$ contains $a_1, a'_1$ since these are the only neighbours
of $a$ in $G_f$.

\setcounter{claim}{0}

\begin{claim}\label{clm:zgxi}
For $i=1, \ldots, n$, $Z$ contains exactly $6m+10$ nodes of $G(z_i)$: four
of these are $a_i, a'_i, b_i, b'_i$, and the others are either the
$t_{i, q}, t'_{i, q}$'s or the $f_{i, q}, f'_{i,q}$'s where $q \in
\{0, \ldots, 3m+2\}$.
\end{claim}
\emph{Proof.} First we prove the claim for $i=1$.  Since $a, a_1$ are
in $Z$ and $a_1$ has only three neighbours $a, t_{1, 0}, f_{1, 0}$,
exactly one of $t_{1, 0}, f_{1, 0}$ is in $Z$.  Likewise exactly one
of $t'_{1, 0}, f'_{1, 0}$ is in $Z$.  If $t_{1, 0}, f'_{1, 0}$ are in
$Z$ then the nodes $a, a_1, a'_1, t_{1, 0}, f'_{1, 0}$ are all in $Z$
and they induce a hole that does not contain $b$, a contradiction.
Likewise we do not have both $t'_{1, 0}, f_{1,0}$ in $Z$.  Therefore,
up to symmetry we may assume that $t_{1, 0}, t'_{1, 0}$ are in $Z$ and
$f_{1, 0}, f'_{1, 0}$ are not. This implies $t_{1, 1}, t'_{1, 1} \in
Z$. 

Suppose that for some $j \in \{1, \ldots, m+1\}$ and $k \in \{0, 1,
2\}$ one of $t_{1, 3(j-1)+k}$, $t'_{1, 3(j-1)+k}$ is not in $Z$.  Let
$3(j-1)+k$ be minimum with that property and assume up to a symmetry
that $t_{1, 3(j-1)+k}$ is not in $Z$.  If $k=0$ or $k=1$, then $Z$
contains up to a symmetry $f_{1, 3(j-1)+k}$ that is adjacent to
$t'_{1, 3(j-1)+k}$, so $Z$ cannot contain $a, b$, a contradiction.
So, $k=2$, $t_{1, 3(j-1)+k}$ is in $Z$ and one of $v_j^p, w_j^p$, $p\in
\{1, 2, 3\}$, say $v_j^1$ up to a symmetry, must be in $Z$.  But then,
by the definition of $G_f$, $v_j^1$ is adjacent to $t'_{1, 3(j-1)+2}$.  So,
$Z$ has node set $\{a, t_{1, 0}, t'_{1, 0}, \ldots, t_{1, 3(j-1)+2}, t'_{1,
  3(j-1)+2}, v_j^1\}$ and $b$ is not in $Z$, a contradiction. So, for $q=
0, \dots, 3m+2$, we have $t_{1, j}, t'_{1, j} \in Z$ and $b_1, b'_1
\in Z$.

Now, for $j= 0, \dots, m$, none of $f_{i, 3j}$, $f_{i, 3j+1}$ can be
in $Z$. So, $f_{i, 3j+2}$ cannot be in $Z$ because only one of its
neighbor can be in $Z$.  In particular, $a_2 \in Z$ and similarly
$a'_2 \in Z$.

 This proves our claim for $i=1$. The proof of the claim for $i=2$ is
 essentially the same as for $i=1$, and by induction the claim holds
 up to $i=n$.  $\Box$

\begin{claim}\label{clm:zgcj}
 For $j=1, \ldots, m$, $Z$ contains $c_j, d_j$ and exactly one of
 $\{u_j^1, v_j^1,$ $w_j^1, x_j^1\}$, $\{u_j^2, v_j^2,$ $w_j^2,
 x_j^2\}$, $\{u_j^3, v_j^3,$ $w_j^3, x_j^3\}$.
\end{claim}
\emph{Proof.} First we prove this claim for $j=1$.  By
Claim~\ref{clm:zgxi}, $b'_n$ is in $Z$ and exactly one of $t'_{n, 3m+2},
f'_{n, 3m+2}$ is in $Z$, so (since $b'_n$ has degree $3$ in $G_f$) $c_1$
is in $Z$.  Consequently exactly one of $u_1^1, u_1^2, u_1^3$ is in
$Z$, say $u_1^1$.  The neighbour of $u_1^1$ in $Z\setminus c_1$ cannot
be a node among $u_1^2, u_1^3$ for this would imply $Z$ that contains
a triangle.  Hence $v_1^1 \in Z$.  The neighbour of $v_1^1$ in
$Z\setminus u_1^1$ cannot be in some $G(z_i)$ ($1\le i\le n$), for in
that case that neighbour would be either $t_{i, 2}$ (or $f_{i, 2}$)
and thus, by Claim~\ref{clm:zgxi}, node $t'_{i, 2}$ (or $f'_{i, 2}$)
would be a third neighbour of $v_1^1$ in $Z$, a contradiction.  Thus
the other neighbour of $v_1^1$ in $Z$ is $w_1^1$.  Similarly, we prove
that $w_1^1, x_1^1, d_1$ are in $Z$, and so the claim holds for $j=1$.
Since $d_1$ has degree $4$ in $G_f$ and exactly one of $x_1^1, x_1^2,
x_1^3$ is in $Z$, it follows that the fourth neighbour $c_2$ of $d_1$
is in $Z$.  Now the proof of the claim for $j=2$ is the same as for
$j=1$, and by induction the claim holds up to $j=m$.  $\Box$

We can now make a Boolean vector $\xi$ as follows.  For $i=1, \ldots,
n$, if $Z$ contains $t_{i, 0}, t'_{i, 0}$ set $\xi_i = 1$; if $Z$
contains $f_{i, 0}, f'_{i, 0}$ set $\xi_i = 0$.  By
Claim~\ref{clm:zgxi} this is consistent.  Consider any clause $C_j$
($1\le j\le m$).  By Claim~\ref{clm:zgcj} and up to symmetry we may
assume that $v_j^1$ is in $Z$.  If $y_j^1 = z_i$ for some $i\in\{1,
.., n\}$, then the construction of $G_f$ implies that $f_{i, 3j-1},
f'_{i, 3j-1}$ are not in $Z$, so $t_{i, 3j-1}, t'_{i, 3j-1}$ are in
$Z$, so $\xi_i=1$, so clause $C_j$ is satisfied by $x_i$.  If $y_j^1 =
\overline{z}_i$ for some $i\in\{1, .., n\}$, then the construction of
$G_f$ implies that $t_{i, 3j-1}, t'_{i, 3j-1}$ are not in $Z$, so
$f_{i, 3j-1}, f'_{i, 3j-1}$ are in $Z$, so $\xi_i=0$, so clause $C_j$
is satisfied by $\overline{z}_i$.  Thus $\xi$ is a truth assignment
for $f$.  This completes the proof of the lemma.  $\Box$

\begin{theorem}
  \label{thm:npc}
  For each integer $k\geq 4$, the problem of deciding whether a graph
  contains a $kPC(\cdot, \cdot)$ is NP-complete, and the problem of
  deciding whether a graph contains a $kPC(\Delta, \cdot)$ is
  NP-complete.
\end{theorem}

\emph{Proof.} Let $k\geq 4$ be an integer.  We give a reduction from
problem $\Pi$ to the problems whose NP-completeness is claimed.  So
let $(G, a, b)$ be any instance of problem $\Pi$, where $G$ is a
$K_{1, 4}$-free graph and $a, b$ are non-adjacent nodes of $G$ of
degree $2$.  Let us call $a', a''$ the two neighbours of $a$ and $b',
b''$ the two neighbours of $b$ in $G$.

\noindent\emph{Reduction to the detection of a $kPC(\cdot,\cdot)$:}
Starting from $G$, build a graph $G'$ as follows: Add nodes $y_1,
\ldots, y_{k-2}$.  Add edges $ay_1$, \ldots, $ay_{k-2}$, $by_1$, \ldots,
$by_{k-2}$.  We see that $G'$ contains a $kPC(\cdot,\cdot)$ if and
only if $G$ contains a hole that contains $a$ and $b$.  So every
instance of $\Pi$ can be reduced polynomially to an instance of the
detection of a $kPC(\cdot,\cdot)$, which proves that this problem is
NP-complete.

\noindent\emph{Reduction to the detection of a $kPC(\Delta, \cdot)$:}
Starting from $G$, build the same graph $G'$ as above.  Subdivide
every edge $ay_i$, $i\in {1, \ldots, k-2}$, by adding a node $z_i$ of
degree~2.  Subdivide edge $aa'$ by adding a node $a'''$ of degree~2.
Add every possible edge between the nodes of $\{b', b'', y_1, \ldots,
y_{k-2}\}$.  We see that $G'$ contains a $kPC(\Delta, \cdot)$ if and
only if $G$ contains a hole that contains $a$ and $b$.  So every
instance of $\Pi$ can be reduced polynomially to an instance of the
detection of a $kPC(\Delta,\cdot)$, which proves that this problem is
NP-complete.  \qed

\end{document}